\input harvmac
\input epsf

\noblackbox

\newcount\figno

\figno=0
\def\fig#1#2#3{
\par\begingroup\parindent=0pt\leftskip=1cm\rightskip=1cm\parindent=0pt
\baselineskip=11pt
\global\advance\figno by 1
\midinsert
\epsfxsize=#3
\centerline{\epsfbox{#2}}
\vskip 12pt
\centerline{{\bf Figure \the\figno} #1}\par
\endinsert\endgroup\par}
\def\figlabel#1{\xdef#1{\the\figno}}
\def\pano{\par\noindent}

\font\cmss=cmss10
\font\cmsss=cmss10 at 7pt

\def\rlx{\relax\leavevmode}
\def\inbar{\vrule height1.5ex width.4pt depth0pt}
\def\IC{\relax\,\hbox{$\inbar\kern-.3em{\rm C}$}}
\def\IR{\relax{\rm I\kern-.18em R}}
\def\IN{\relax{\rm I\kern-.18em N}}
\def\IP{\relax{\rm I\kern-.18em P}}
\def\ZZ{\rlx\leavevmode\ifmmode\mathchoice{\hbox{\cmss Z\kern-.4em Z}}
 {\hbox{\cmss Z\kern-.4em Z}}{\lower.9pt\hbox{\cmsss Z\kern-.36em Z}}
 {\lower1.2pt\hbox{\cmsss Z\kern-.36em Z}}\else{\cmss Z\kern-.4em Z}\fi}

\def\narrowplus{\kern -.04truein + \kern -.03truein}
\def\narrowminus{- \kern -.04truein}
\def\narrowminussub{\kern -.02truein - \kern -.01truein}

\def\o#1{\overline{#1}}



\lref\KasteID{
P.~Kaste, W.~Lerche, C.~A.~Lutken and J.~Walcher,
{\it D-branes on K3-fibrations},
Nucl.\ Phys.\ B {\bf 582} (2000) 203,
hep-th/9912147.
}

\lref\rblumb{J.~D.~Blum, {\it F Theory Orientifolds, M Theory Orientifolds and
Twisted Strings}, Nucl.Phys. B {\bf 486} (1997) 34, hep-th/9608053.
}

\lref\rcveticb{M.~Cvetic, G.~Shiu and  A.~M.~Uranga,  {\it
Chiral Four-Dimensional N=1 Supersymmetric Type IIA Orientifolds from
Intersecting D6-Branes}, Nucl. Phys. B {\bf 615} (2001) 3, hep-th/0107166.
}

\lref\DabholkarKA{
A.~Dabholkar and J.~Park,
{\it A Note on Orientifolds and F-theory},
Phys.\ Lett.\ B {\bf 394} (1997) 302, hep-th/9607041.
}

\lref\refPradisi{
G.~Pradisi,
{\it Type I Vacua from Diagonal $\ZZ_3$-Orbifolds}, 
Nucl.\ Phys.\ B {\bf 575} (2000) 134,
hep-th/9912218.
}

\lref\BurgessPX{
C.~P.~Burgess, L.~E.~Ibanez and F.~Quevedo,
{\it Strings at the intermediate scale or is the 
Fermi scale dual to the  Planck scale?},
Phys.\ Lett.\ B {\bf 447} (1999) 257, 
hep-ph/9810535.
}

\lref\HananyPY{
A.~Hanany and A.~Iqbal,
{\it Quiver Theories from D6-branes via Mirror Symmetry},
JHEP {\bf 0204} (2002) 009, 
hep-th/0108137.
}

\lref\FengBN{
B.~Feng, A.~Hanany, Y.~H.~He and A.~M.~Uranga,
{\it Toric Duality as Seiberg Duality and Brane Diamonds},
JHEP {\bf 0112} (2001) 035, 
hep-th/0109063.
}

\lref\RandallVF{
L.~Randall and R.~Sundrum,
{\it An Alternative to Compactification}, 
Phys.\ Rev.\ Lett.\  {\bf 83} (1999) 4690, 
hep-th/9906064.
}

\lref\RandallEE{
L.~Randall and R.~Sundrum,
{\it A Large Mass Hierarchy from a Small Extra Dimension}, 
Phys.\ Rev.\ Lett.\  {\bf 83} (1999) 3370, 
hep-ph/9905221.
}

\lref\AngelantonjCT{
C.~Angelantonj and A.~Sagnotti,
{\it Open Strings},
hep-th/0204089.
}


\lref\rangles{M.~Berkooz, M.~R.~Douglas and R.~G.~Leigh, {\it Branes Intersecting
at Angles}, Nucl. Phys. B {\bf 480} (1996) 265, hep-th/9606139.
}

\lref\KleinVU{
M.~Klein,
{\it Couplings in Pseudo-Supersymmetry}, hep-th/0205300.
}

\lref\KleinJR{
M.~Klein,
{\it Loop-Effects in Pseudo-Supersymmetry}, 
hep-th/0209206.
}

\lref\berlin{R.~Blumenhagen, B.~K\"ors and D.~L\"ust,
{\it Moduli Stabilization for Intersecting Brane Worlds in Type 0$\, '$
String Theory}, 
Phys.\ Lett.\ B {\bf 532} (2002) 141, hep-th/0202024.
}

\lref\rrab{R.~Rabadan, {\it Branes at Angles, Torons, Stability and
Supersymmetry}, Nucl.\ Phys.\ B {\bf 620} (2002) 152, hep-th/0107036.
}

\lref\rbgkb{R.~Blumenhagen, L.~G\"orlich and B.~K\"ors,
{\it Supersymmetric 4D Orientifolds of Type IIA with D6-branes at Angles},
JHEP {\bf 0001} (2000) 040,  hep-th/9912204.
}

\lref\rbgkbsum{R.~Blumenhagen, L.~G\"orlich and B.~K\"ors,
{\it A New Class of Supersymmetric Orientifolds with D-Branes at
Angles}, hep-th/0002146.
}

\lref\rbgklnon{R.~Blumenhagen, L.~G\"orlich, B.~K\"ors and D.~L\"ust,
{\it Noncommutative Compactifications of Type I Strings on Tori with Magnetic
Background Flux}, JHEP {\bf 0010} (2000) 006, hep-th/0007024.
}

\lref\rbgklmag{R.~Blumenhagen, L.~G\"orlich, B.~K\"ors and D.~L\"ust,
{\it Magnetic Flux in Toroidal Type I Compactification}, Fortsch. Phys. 49
(2001) 591, hep-th/0010198.
}

\lref\ras{C.~Angelantonj, A.~Sagnotti, {\it Type I
Vacua and Brane Transmutation}, hep-th/0010279.
}

\lref\raads{C.~Angelantonj, I.~Antoniadis, E.~Dudas, A.~Sagnotti, {\it Type I
Strings on Magnetized Orbifolds and Brane Transmutation},
Phys. Lett. B {\bf 489} (2000) 223, hep-th/0007090.
}

\lref\rbkl{R.~Blumenhagen, B.~K\"ors and D.~L\"ust,
{\it Type I Strings with $F$ and $B$-Flux}, JHEP {\bf 0102} (2001) 030,
hep-th/0012156.
}

\lref\rbgkl{R.~Blumenhagen, L.~G\"orlich, B.~K\"ors and D.~L\"ust,
{\it Asymmetric Orbifolds, Noncommutative Geometry and Type I
Vacua}, Nucl.\ Phys.\ B {\bf 582} (2000) 44, hep-th/0003024.
}

\lref\rbgka{R.~Blumenhagen, L.~G\"orlich and B.~K\"ors,
{\it Supersymmetric Orientifolds in 6D with D-Branes at Angles},
Nucl. Phys. B {\bf 569} (2000) 209, hep-th/9908130.
}

\lref\rcvetica{M.~Cvetic, G.~Shiu and  A.~M.~Uranga,  {\it Three-Family
Supersymmetric Standard-like Models from Intersecting Brane Worlds}, 
Phys. Rev. Lett. {\bf 87} (2001) 201801,  hep-th/0107143.
}

\lref\rott{R.~Blumenhagen, B.~K\"ors, D.~L\"ust and T.~Ott, {\it
The Standard Model from Stable Intersecting Brane World Orbifolds},
Nucl. Phys. B {\bf 616} (2001) 3, hep-th/0107138.
}

\lref\rottb{R.~Blumenhagen, B.~K\"ors, D.~L\"ust and T.~Ott, 
{\it Intersecting Brane Worlds on Tori and Orbifolds}, 
Fortsch.\ Phys.\  {\bf 50} (2002) 843, hep-th/0112015.
}

\lref\rbonna{S.~F\"orste, G.~Honecker and R.~Schreyer, {\it
Orientifolds with Branes at Angles}, JHEP {\bf 0106} (2001) 004,
hep-th/0105208.
}

\lref\rbonnb{G.~Honecker, {\it Intersecting Brane World Models from
D8-branes on $(T^2 \times T^4/\ZZ_3)/\Omega R_1$ Type IIA Orientifolds},
JHEP {\bf 0201} (2002) 025, hep-th/0201037.
}

\lref\refHonecker{
S.~F\"orste, G.~Honecker and R.~Schreyer, 
{\it Supersymmetric $Z_N \times Z_M$ Orientifolds in 4D with D-Branes at Angles}, 
Nucl.\ Phys.\ B {\bf 593} (2001) 127,
hep-th/0008250.
}

\lref\rqsusy{D.~Cremades, L.~E.~Ibanez and F.~Marchesano, {\it
SUSY Quivers, Intersecting Branes and the Modest Hierarchy Problem},
JHEP {\bf 0207} (2002) 009, 
hep-th/0201205.
}

\lref\rqsusyb{D.~Cremades, L.~E.~Ibanez and F.~Marchesano, {\it
     Intersecting Brane Models of Particle Physics and the Higgs Mechanism},
JHEP {\bf 0207} (2002) 022,     
hep-th/0203160.
}

\lref\rbachas{C.~Bachas, {\it A Way to Break Supersymmetry}, hep-th/9503030.
}

\lref\rafiruph{G.~Aldazabal, S.~Franco, L.~E.~Ibanez, R.~Rabadan, A.~M.~Uranga,
{\it Intersecting Brane Worlds}, JHEP {\bf 0102} (2001) 047, hep-ph/0011132.
}

\lref\rafiru{G.~Aldazabal, S.~Franco, L.~E.~Ibanez, R.~Rabadan, A.~M.~Uranga,
{\it $D=4$ Chiral String Compactifications from Intersecting Branes}, 
J.\ Math.\ Phys.\  {\bf 42} (2001) 3103, hep-th/0011073.
}

\lref\rimr{L.~E.~Ibanez, F.~Marchesano, R.~Rabadan, {\it Getting just the
Standard Model at Intersecting Branes}, 
JHEP {\bf 0111} (2001) 002, hep-th/0105155.
}

\lref\belrab{J.~Garcia-Bellido and R.~Rabadan, {\it Complex Structure Moduli
Stability in Toroidal Compactifications}, 
JHEP {\bf 0205} (2002) 042, hep-th/0203247.
}

\lref\rcim{D.~Cremades, L.~E.~Ibanez and F.~Marchesano, {\it
    Standard Model at Intersecting D5-branes: Lowering the String Scale},
     hep-th/0205074.
}

\lref\rkokoa{
C.~Kokorelis, 
{\it GUT Model Hierarchies from Intersecting Branes},
JHEP {\bf 0208} (2002) 018, hep-th/0203187.
}

\lref\EllisKantiNano{
J.~R.~Ellis, P.~Kanti and D.~V.~Nanopoulos,
{\it Intersecting branes flip SU(5)},
hep-th/0206087.
}

\lref\rkokob{C.~Kokorelis, {\it New Standard Model Vacua from Intersecting Branes},
JHEP {\bf 0209} (2002) 029, hep-th/0205147.
}

\lref\BlumenhagenWN{
R.~Blumenhagen, V.~Braun, B.~K\"ors and D.~L\"ust,
{\it Orientifolds of K3 and Calabi-Yau Manifolds with Intersecting D-branes}, 
JHEP {\bf 0207} (2002) 026, 
hep-th/0206038.
}

\lref\radd{N.~Arkani-Hamed, S.~Dimopoulos, and G.~Dvali, {\it The Hierarchy
Problem and New Dimensions at a Millimeter}, Phys. Lett. B {\bf 429} (1998)
263, hep-ph/9803315.
}

\lref\raadd{I.~Antoniadis, N.~Arkani-Hamed, S.~Dimopoulos, and G.~Dvali, {\it
New Dimensions at a Millimeter to a Fermi and Superstrings at a TeV},
Phys. Lett. B {\bf 436} (1998) 257, hep-ph/9804398.
}

\lref\rscruccab{C.~A.~Scrucca and M.~Serone,
   {\it Anomalies and Inflow on D-branes and O-planes},
   Nucl. Phys. B {\bf 556} (1999) 197, hep-th/9903145.
}

\lref\rscrucca{J.~F.~Morales, C.~A.~Scrucca and M.~Serone,
        {\it Anomalous Couplings for D-branes and O-planes},
      Nucl. Phys. B {\bf 552} (1999) 291, hep-th/9812071.
}


\lref\BailinIE{
D.~Bailin, G.~V.~Kraniotis and A.~Love,
{\it Standard-like Models from Intersecting D4-branes},
Phys.\ Lett.\ B {\bf 530} (2002) 202,
hep-th/0108131.
}


\lref\sagn{M.~Bianchi and A.~Sagnotti, {\it On the Systematics of Open String
Theories}, Phys. Lett. B {\bf 247} (1990) 517.
}

\lref\rgimpol{
E.~G.~Gimon and J.~Polchinski, {\it Consistency Conditions
for Orientifolds and D-Manifolds}, Phys.\ Rev.\ D {\bf 54} (1996) 1667,
hep-th/9601038.
}

\lref\rgimjo{ E.~G.~Gimon and C.~V.~Johnson, {\it K3 Orientifolds},
Nucl.\ Phys.\ B {\bf 477} (1996) 715, hep-th/9604129.
}

\lref\rbluma{J.~D.~Blum and A.~Zaffaroni, {\it An Orientifold from F Theory},
Phys.\ Lett.\ B {\bf 387} (1996) 71, hep-th/9607019.
}

\lref\rdabol{ A.~Dabholkar and J.~Park, {\it An Orientifold of Type
        IIB theory on K3}, Nucl. Phys. B {\bf 472} (1996) 207,
        hep-th/9602030;
        {\it Strings on Orientifolds}, Nucl. Phys. B {\bf 477}
        (1996) 701, hep-th/9604178.
}


\lref\rbdlr{I.~Brunner, M.~R.~Douglas, A.~Lawrence and C.~R\"omelsberger, {\it
  D-branes on the Quintic}, JHEP 0008 (2000) 015, hep-th/9906200.
}

\lref\rhiv{K.~Hori, A.~Iqbal and C.~Vafa, {\it
            D-Branes And Mirror Symmetry}, hep-th/0005247.
}

\lref\rkklma{S.~Kachru, S.~Katz, A.~Lawrence and J.~McGreevy, {\it
      Open String Instantons and Superpotentials}, Phys. Rev. D {\bf 62} (2000)
      026001, hep-th/9912151.
}

\lref\rmayr{P.~Mayr, {\it N=1 Mirror Symmetry and Open/Closed String Duality},
 hep-th/0108229.
}

\lref\PlebanskiGY{
J.~F.~Plebanski and M.~Demianski,
{\it Rotating, Charged, and Uniformly Accelerating Mass in General 
Relativity},
Annals Phys.\  {\bf 98} (1976) 98.
}


\lref\rghm{M.~Green, J.~A.~Harvey, G.~Moore, 
{\it I-Brane Inflow and Anomalous Couplings on D-Branes}, 
Class. Quant. Grav. {\bf 14} (1997) 47, hep-th/9605033.
}

\lref\DouglasBN{
M.~R.~Douglas,
{\it Branes within Branes}, hep-th/9512077.
}


\lref\AtiyahZZ{
M.~Atiyah, J.~M.~Maldacena and C.~Vafa,
{\it An M-theory Flop as a Large N Duality},
J.\ Math.\ Phys.\  {\bf 42} (2001) 3209, hep-th/0011256.
}

\lref\BrandhuberYI{
A.~Brandhuber, J.~Gomis, S.~S.~Gubser and S.~Gukov,
{\it Gauge Theory at Large N and New $G_2$ Holonomy Metrics},
Nucl.\ Phys.\ B {\bf 611} (2001) 179, hep-th/0106034.
}

\lref\CveticZX{
M.~Cvetic, G.~W.~Gibbons, H.~Lu and C.~N.~Pope,
{\it Cohomogeneity One Manifolds of Spin(7) and $G_2$ Holonomy},
Phys.\ Rev.\ D {\bf 65} (2002) 106004, hep-th/0108245.
}

\lref\rbs{
R.~L.~Bryant and S.~Salamon, {\it On the Construction of some Complete
Metrics with Exceptional Holonomy},
Duke Math. J. {\bf 58} (1989) 829.
}

\lref\rberbra{
P.~Berglund and A.~Brandhuber, {\it Matter from $G_2$ Manifolds},
Nucl.\ Phys.\ B {\bf 641} (2002) 351, 
hep-th/0205184.
}

\lref\PapadopoulosDA{
G.~Papadopoulos and P.~K.~Townsend,
{\it Compactification of D = 11 Supergravity on Spaces of Exceptional Holonomy},
Phys.\ Lett.\ B {\bf 357} (1995) 300, hep-th/9506150.
}

\lref\AcharyaGB{
B.~S.~Acharya,
{\it On Realising N = 1 Super Yang-Mills in M theory},
hep-th/0011089.
}

\lref\BrandhuberKQ{
A.~Brandhuber,
{\it $G_2$ Holonomy Spaces from Invariant Three-forms},
Nucl.\ Phys.\ B {\bf 629} (2002) 393, hep-th/0112113.
}

\lref\AnguelovaDD{
L.~Anguelova and C.~I.~Lazaroiu,
{\it M-theory on 'Toric' $G(2)$ Cones and its Type II Reduction},
hep-th/0205070.
}

\lref\AtiyahQF{
M.~Atiyah and E.~Witten,
{\it M-theory Dynamics on a Manifold of $G_2$ Holonomy},
hep-th/0107177.
}

\lref\EguchiIP{
T.~Eguchi and Y.~Sugawara,
{\it String Theory on $G_2$ Manifolds based on Gepner Construction},
Nucl.\ Phys.\ B {\bf 630} (2002) 132, hep-th/0111012.
}

\lref\RoibanCP{
R.~Roiban and J.~Walcher,
{\it Rational Conformal Field Theories with $G_2$ Holonomy},
JHEP {\bf 0112} (2001) 008, hep-th/0110302.
}

\lref\BlumenhagenJB{
R.~Blumenhagen and V.~Braun,
{\it Superconformal Field Theories for Compact $G_2$ Manifolds},
JHEP {\bf 0112} (2001) 006, hep-th/0110232.
}

\lref\rwitten{E.~Witten, {\it Anomaly Cancellation on Manifolds of $G_2$
Holonomy}, hep-th/0108165.
}

\lref\rwa{B.~Acharya and E.~Witten, {\it Chiral Fermions from Manifolds of
$G_2$ Holonomy}, hep-th/0109152.
}

\lref\GibbonsER{
G.~W.~Gibbons, D.~N.~Page and C.~N.~Pope,
{\it Einstein Metrics On $S^3$,  $R^3$ And $R^4$ Bundles},
Commun.\ Math.\ Phys.\  {\bf 127} (1990) 529.
}

\lref\BehrndtYE{
K.~Behrndt,
{\it Singular 7-manifolds with G(2) holonomy and intersecting 6-branes},
Nucl.\ Phys.\ B {\bf 635} (2002) 158, 
hep-th/0204061.
}

\lref\BehrndtXM{
K.~Behrndt, G.~Dall'Agata, D.~L\"ust and S.~Mahapatra,
{\it Intersecting 6-branes from New 7-manifolds with $G(2)$ Holonomy},
JHEP {\bf 0208} (2002) 027, hep-th/0207117.
}

\lref\BlumenhagenUA{
R.~Blumenhagen, B.~K\"ors, D.~L\"ust and T.~Ott,
{\it Hybrid Inflation in Intersecting Brane Worlds},
Nucl.\ Phys.\ B {\bf 641} (2002) 235, 
hep-th/0202124.
}

\lref\VafaWI{
C.~Vafa,
{\it Superstrings and Topological Strings at Large $N$},
J.\ Math.\ Phys.\  {\bf 42} (2001) 2798,
hep-th/0008142.
}

\lref\LercheYW{
W.~Lerche, P.~Mayr and N.~Warner,
{\it N=1 Special Geometry, Mixed Hodge Variations and Toric Geometry},
hep-th/0208039.
}

\lref\LercheCK{
W.~Lerche, P.~Mayr and N.~Warner,
{\it Holomorphic N = 1 Special Geometry of Open-closed Type II Strings},
hep-th/0207259.
}

\lref\LercheCW{
W.~Lerche and P.~Mayr,
{\it On N = 1 Mirror Symmetry for Open Type II Strings},
hep-th/0111113.
}

\lref\CurioDZ{
G.~Curio, B.~K\"ors and D.~L\"ust,
{\it Fluxes and Branes in Type II Vacua and
M-theory Geometry with $G_2$ and  $Spin(7)$ Holonomy},
Nucl.\ Phys.\ B {\bf 636} (2002) 197, 
hep-th/0111165.
}

\lref\refTamarFriedmann{
T.~Friedmann,
{\it On the Quantum Moduli Space of M Theory Compactifications}, 
Nucl.\ Phys.\ B {\bf 635} (2002) 384,
hep-th/0203256.
}

\lref\UrangaPG{
A.~M.~Uranga,
{\it Local Models for Intersecting Brane Worlds}, 
hep-th/0208014.
}


\lref\Stefanski{
B.~Stefa\'nski,~jr,
{\it Gravitational Couplings of D-branes and O-planes}, 
Nucl.\ Phys.\ B {\bf 548} (1999) 275,
hep-th/9812088.
}

\lref\CveticYA{
M.~Cvetic, G.~W.~Gibbons, H.~Lu and C.~N.~Pope,
{\it M3-branes, G(2) Manifolds and Pseudo-Supersymmetry},
Nucl.\ Phys.\ B {\bf 620} (2002) 3, 
hep-th/0106026.
}

\Title{\vbox{
 \hbox{HU--EP-02/44}
 \hbox{SPIN-02/29}
 \hbox{ITP-UU-02/47}
 \hbox{LPTENS-02/52}
 \hbox{hep-th/0210083}
 \vskip -5mm
}}
{\vbox{\centerline{The Standard Model on the Quintic}}
}
\vskip -5mm
\centerline{Ralph Blumenhagen{$^1$}, Volker Braun{$^{1,2}$},
Boris K\"ors{$^3$}, and Dieter L\"ust{$^1$} }
\bigskip
\centerline{$^1$ {\it Humboldt-Universit\"at zu Berlin, Institut f\"ur
Physik,}}
\centerline{\it Invalidenstrasse 110, 10115 Berlin, Germany}
\centerline{\tt e-mail:
blumenha, braun, luest@physik.hu-berlin.de}
\bigskip
\line{
  \hbox to 65mm{\vbox{
    \hbox to 65mm{\hss $^2$ {\it Ecole Normale Sup\'erieure,}\hss}
    \hbox to 65mm{\hss \it 24 rue Lhomond, 75231 Paris, France\hss}
    \hbox to 65mm{\hss \tt email: volker.braun@lpt.ens.fr\hss}
  }}
  \hfill
  \hbox to 65mm{\vbox{
    \hbox to 65mm{\hss $^3$ {\it Spinoza Institute, Utrecht University,}\hss}
    \hbox to 65mm{\hss \it Utrecht, The Netherlands\hss}
    \hbox to 65mm{\hss \tt email: kors@phys.uu.nl\hss}
  }}
}
\bigskip\bigskip

\centerline{\bf Abstract}
\noindent
We describe the general geometrical framework of brane world constructions 
in orientifolds of type IIA string theory with D6-branes 
wrapping 3-cycles
in a Calabi-Yau 3-fold, and point out their immediate 
phenomenological relevance. These branes generically intersect 
in points, and the patterns of intersections govern the chiral fermion 
spectra and issues of gauge and supersymmetry breaking in the low energy 
effective gauge theory on their world volume. 
In particular, we provide an example of an intersecting brane world 
scenario on the quintic Calabi-Yau with the gauge group and the chiral spectrum 
of the Standard Model and discuss its properties in some detail. 
Additionally we explain related technical advancements in the 
construction of supersymmetric orientifold vacua with intersecting D-branes. 
Six-dimensional orientifolds of this type generalize the 
rather limited set of formerly known orbifolds of type I, and the presented 
techniques provide a short-cut to obtain their spectra. 
Finally, we comment on lifting configurations of intersecting D6-branes 
to M-theory on non-compact $G_2$ manifolds. 

\smallskip

\Date{10/2002}

\newsec{Introduction}

A central object of string phenomenology is to find a string vacuum
whose low energy approximation is reproducing the known physics of the
Standard Model or of its supersymmetric and grand unified
extensions. As a first approach one may concentrate on finding models
with the correct light degrees of freedom, the right gauge group and
chiral fermion spectra, leaving the details of their dynamics aside
for the moment. Intersecting brane worlds \rbgklnon\ have proven to be
a candidate framework of model building which offers excellent
opportunity to meet this requirement \refs{\raads\rbgklmag\ras
\rafiru\rafiruph\rbkl\rimr\rbonna
\rott\rcveticb\BailinIE\rbonnb\EllisKantiNano\rkokoa\rcim-\rkokob}.
In these string compactifications, the fermion spectrum is determined
by the intersection numbers of certain 3-cycles in the internal space,
as opposed for instance to the older approaches involving heterotic
strings, where the number of generations was given by the Euler
characteristic in the simplest case. Beyond these topological data
also some more geometrical issues have been addressed, which provide
access to computing the scalar potential and determining the dynamics
at least at the classical level \refs{\rott,\rottb,\berlin}.  Up to
now, the construction has been limited to toroidal backgrounds and
orbifolds thereof for the sake of simplicity. This has actually put
severe constraints on the generality of the examples obtained and
prevented any supersymmetric models, except for the one case of the
$T^6/\ZZ^2_2$ Calabi-Yau-orbifold \refs{\rcvetica\rcveticb},
generalizing the former work on the six-dimensional $T^4/\ZZ_2$
K3-orbifold \rbkl.

In the present work, which covers only a fraction of the material
presented first in \BlumenhagenWN, but also adds some extensions and
slight improvements, we mostly describe the general framework of
intersecting brane world constructions on any smooth background
Calabi-Yau space\footnote{$^1$}{See also the recent work \UrangaPG\
where non-compact local models of intersection brane configurations on
Calabi-Yau spaces have been discussed.}.  This extends the range of
accessible background spaces to include basically all geometrical
string compactifications with supersymmetry in the bulk gravity
sector. Therefore obstacles to finding supersymmetric Standard Model
or GUT compactifications may possibly be overcome. For the time being, 
we give a new solution for a non-supersymmetric intersecting brane world 
on the quintic Calabi-Yau that not only realizes the chiral fermion 
spectrum of the Standard Model but can also be shown to have exactly 
the hypercharge as its only abelian gauge symmetry. It replaces the 
example given in \BlumenhagenWN. The general
philosophy of these models has been sketched in figure 1.

\fig{}{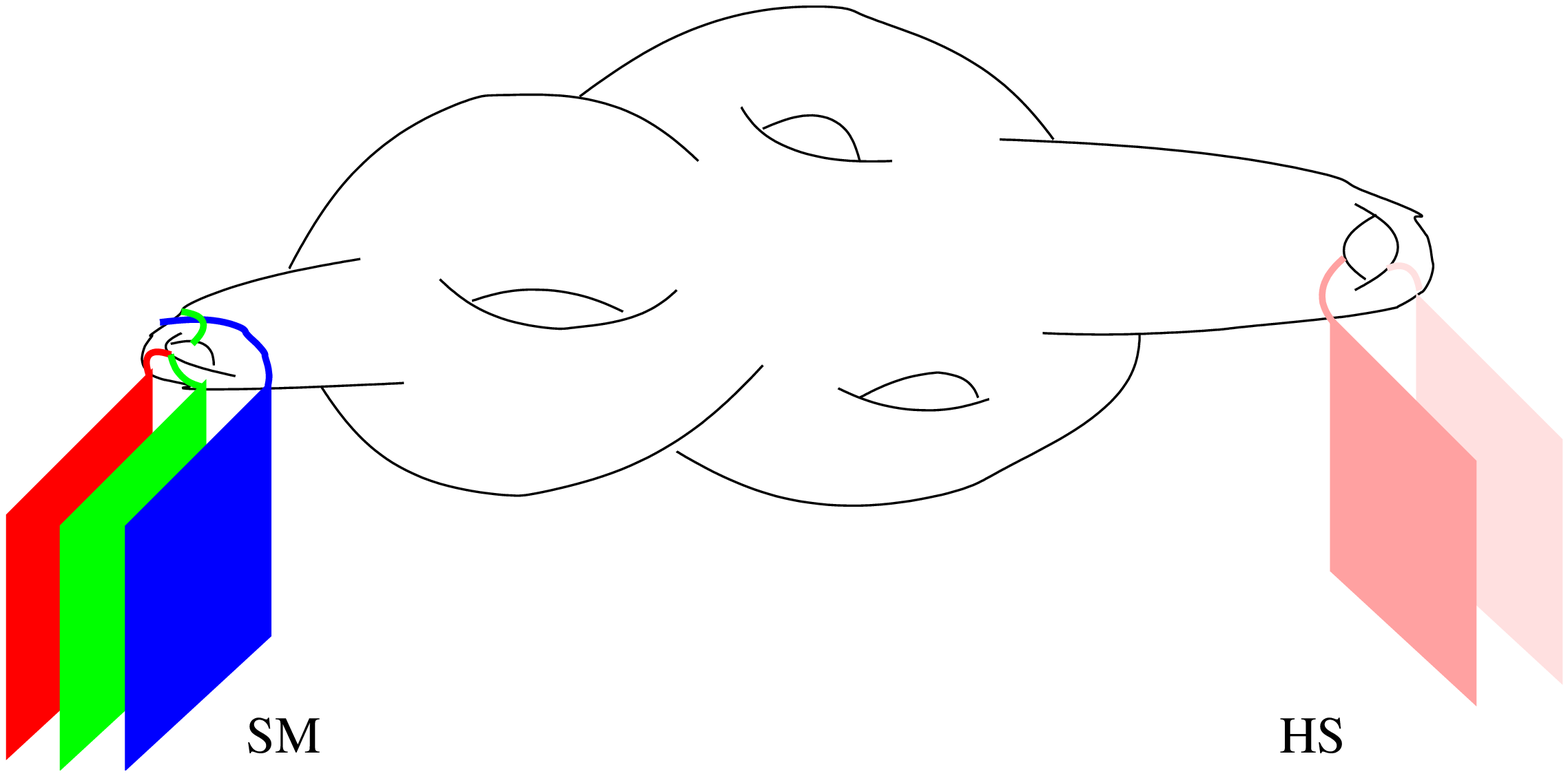}{12truecm} 

The internal Calabi-Yau space may split into various pieces which
individually support 3-cycles wrapped by several D6-branes. On one of
these regions, the Standard Model (SM) fields are localized, while
others may involve hidden sector (HS) gauge groups which couple only
gravitationally to the visible sector (see figure 1).  This kind of
scenario offers at least two possible ways to address the issue of
space-time supersymmetry breaking in intersecting brane worlds.  In
the first class of models the Standard Model brane configuration is
already non-supersymmetric from the beginning (this is true for most
of the models considered so far, including the CY example in
\BlumenhagenWN).  This means that supersymmetry is broken at the
string scale $M_{\rm string}$.  In order to avoid the usual hierarchy
problems $M_{\rm string}$ should be of the order of a few TeV,
requiring that the volume of the internal CY space transverse to all
Standard Model D-branes is large according to \refs{\radd,\raadd}.

On the other hand it may happen that the Standard Model D-brane sector
is itself supersymmetric, but the hidden sector preserves a different
supersymmetry\footnote{$^2$}{The relevant patterns of supersymmetry
breaking in the effective low energy field theory have been discussed
in \refs{\rqsusy,\KleinVU,\KleinJR}.} or is completely
non-supersymmetric. Then the gravity mediated supersymmetry breaking
appears naturally. In this case the following relation between the
SUSY breaking scale in the Standard Model sector and the fundamental
string scale is expected to hold:
\eqn\susy
{M_{3/2}\simeq{M_{\rm string}^2\over M_{\rm Planck}}\, .
}
With $M_{3/2}$ of order TeV one obtains an intermediate string scale,
$M_{\rm string}\simeq 10^{11}{\rm GeV}$, 
a scenario which was already discussed in \BurgessPX.
For
D6-brane models the  string scale, 
the string coupling
constant $g_{\rm string}$, the typical length scale $R_\parallel$ 
of the internal D6-brane volume Vol$({\rm D}6) \sim R_\parallel^3$ and
the scale $R_\perp$ of the transversal internal volume are related to the
gauge coupling $g_{\rm YM}$ and the effective Planck mass in the following way:
\eqn\radii
{g^2_{\rm YM}=g_{\rm string}(M_{\rm string}R_\parallel)^{-3}\, ,
\qquad
M_{\rm Planck}={M_{\rm string}^4\over g_{\rm string}}(R_\parallel
R_\perp)^{3/2}\, .}
Assuming that $M_{\rm string}\simeq R_\parallel^{-1}$
this requires a moderately enlarged transversal space, namely
$R^{-1}_\perp\simeq 10^9{\rm GeV}$.

Finally there is another effect known to weaken the breaking effects in the 
effective 
four-dimensional theory on the Standard Model branes and to avoid the 
hierarchy problems.
Namely, the backreaction of the geometry towards the presence of 
the branes, which is actually neglected in our approach, may involve a warped 
geometry, which may give rise to a scenario in the spirit of 
\refs{\RandallEE,\RandallVF}. 

As an additional important motivation, we also hope to gain further
insight into the dynamics of the models from the studies of Calabi-Yau
geometry and mirror symmetry, in particular from the knowledge of
scalar potentials
\refs{\rbdlr\KasteID\rkklma\rhiv\HananyPY\rmayr\FengBN\LercheCW\LercheCK-\LercheYW}.
For example, it was recently pointed out that type IIB vacua deformed
by 3-form fluxes and D5-branes possess a rather particular structure
reminiscent of the special geometry that governs the ${\cal N}=2$
vacuum of the Calabi-Yau compactification \refs{\LercheCK,\LercheYW}.

An essential consistency condition is the cancellation of the
Ramond-Ramond (RR) charge. There are two complementary methods in
performing the actual computations of the cancellation as well as of
the chiral spectra, the scalar potential and other relevant data. If
the background is given by a string world sheet CFT, these can be
extracted from certain string diagrams, notably the genus zero open
string one-loop amplitude.  On the other hand, if the background is
smooth, one may go to the limit where the curvature is small
everywhere and supergravity and classical geometry are valid. Since we
are mostly interested in backgrounds defined in geometrical terms, we
shall employ arguments taken from the effective supergravity action
and from geometry. One of the main achievements of \BlumenhagenWN\
actually was to show that for orbifold vacua, where the two regimes
are connected by blowing up singularities, the geometrical point of
view provides a much simpler formalism to compute the chiral spectra
as compared to the CFT methods. We shall demonstrate this by studying
orbifold limits of K3 in their blown-up version.

\newsec{Intersecting brane worlds on Calabi-Yau 3-folds}
\seclab\sIBWintro

In the brane world scenarios we are going to consider here, there are 
D-branes filling out the entire four-dimensional space-time providing the 
degrees of freedom for an effective gauge theory. The overall 
transverse six-dimensional space is compact, such that the internal 
excitations decouple from the effective theory below the string scale.  
The global consistency conditions in string models with D-branes that fill 
out the non-compact space-time 
involve the cancellation of the RR 
charges. Furthermore, supersymmetry requires the cancellation of the 
brane tensions and the corresponding Neveu-Schwarz-Neveu-Schwarz (NSNS) 
tadpoles as well. If the latter is 
neglected, one can achieve the RR charge cancellation within type II vacua by 
including anti-branes, but these vacua usually suffer from run-away 
instabilities, if not even tachyons. The only setting in which objects with 
negative tension arise naturally in string theory are 
orientifolds \AngelantonjCT, where 
the orientifold O-planes can balance the charge and tension of the D-branes. 
Therefore, orientifolds provide the framework where supersymmetric 
brane worlds may be found within string theory. 

\subsec{Definition}

According to the above reasoning we will consider 
orientifold compactifications, where the ten-dimensional space-time 
${\cal X}$ is of the kind 
\eqn\spacetime{
{\cal X} = \IR^{3,1} \times {{\cal M}^{6} \over \Omega \o\sigma} . 
}
Here ${\cal M}^{6}$ is a Calabi-Yau 3-fold with a symmetry under 
$\o\sigma$, the complex conjugation 
\eqn\sigmadef{
\o\sigma : z_i \mapsto \bar{z}_i,\ i=1,\, ...\, ,3,
}
in local coordinates. It is combined with the world sheet parity $\Omega$ 
to form the orientifold projection $\Omega \o\sigma$. 
This operation is actually a symmetry of the type IIA string 
on ${\cal M}^{6}$. 
The construction has a T-dual or mirror symmetric description within 
type IIB, which is explained in some detail in \BlumenhagenWN\ as well. 
Orientifold O6-planes are defined as the fixed locus Fix$(\o\sigma)$ 
of $\o\sigma$, which is easily seen to be a supersymmetric 3-cycle in 
${\cal M}^{6}$. It is special Lagrangian (sLag) and calibrated with 
respect to the real part of the holomorphic 3-form $\Omega_3$. 
To see this define $\Omega_3$ and the K\"ahler form $J$ in local coordinates 
\eqn\nform{
\Omega_3 = dz_1 \wedge dz_2 \wedge dz_3 , \quad
J = i \sum_{i=1}^3{ dz_i \wedge d\bar{z}_i} .
} 
From $\o\sigma( \Omega_3 ) = \o{\Omega}_3$ and $\o\sigma( J ) = -J$ it
then follows that
\eqn\slagone{
\Im(\Omega_3)\vert_{{\rm Fix}(\o\sigma)} =0  ,\quad
J\vert_{{\rm Fix}(\o\sigma)} =0 ,
}
which implies 
\eqn\slagtwo{
\Re(\Omega_3)\vert_{{\rm Fix}(\o\sigma)} = 
d{\rm vol}\vert_{{\rm Fix}(\o\sigma)} .
}
It is also useful to define a rescaled 3-form 
\eqn\normom{
\widehat \Omega_3={1\over \sqrt{{\rm Vol}({\cal M}^{6})}}\Omega_3 .
}
This orientifold projection truncates the gravitational bulk theory of closed 
strings down to a theory with $16$ supercharges in ten dimensions, 
leading to 4 supercharges and ${\cal N}=1$ in four dimensions, after 
compactifying on the Calabi-Yau. In order to cancel the RR charge of 
the O6-planes it is required to introduce D6-branes into the theory as well, 
which will provide the gauge sector of the theory. If we label the 
individual stacks of D6$_a$-branes with multiplicities $N_a$ 
by a label $a$, the gauge group of the effective theory will be given by 
\eqn\gauge{
G=\prod_{a} U(N_a) .
}
Here we exclude the possibility of branes which are invariant under the 
projection $\Omega\o\sigma$. They would give rise to $SO(N_a)$ or 
$Sp(N_a)$ factors. It is no conceptual problem to include them as well, but 
they are of little phenomenological interest. 

\subsec{RR charges and brane tension}

The charge cancellation conditions are often obtained by regarding 
divergences of one-loop open string amplitudes, but can also be 
determined from the consistency of the background in the 
supergravity equations of motion or Bianchi identities. 
The Chern-Simons action for
D$p$-branes and O$p$-planes are given by 
\refs{\DouglasBN\rghm\rscrucca\rscruccab-\Stefanski}
\eqn\bornina{\eqalign{
{\cal S}^{({\rm D}p)}_{\rm CS} &=
\mu_p \int_{{\rm D}p}
{\rm ch}({\cal F})\wedge  \sqrt{{\hat {\cal A}({\cal R}_T) \over \hat {\cal A}({\cal R}_N)
   }} \wedge \sum_q{C_q} , \cr 
{\cal S}^{({\rm O}p)}_{\rm CS} &=
Q_p \mu_p \int_{{\rm O}p}
    \sqrt{{\hat {\cal L}({\cal R}_T/4) \over \hat{\cal L}({\cal R}_N/4)
   }} \wedge \sum_q{C_q} .
}}
The relative charge of the orientifold planes is given by
$Q_p=-2^{p-4}$ and ch$({\cal F})$ denotes the Chern character, 
$\hat{\cal A}({\cal R})$ the Dirac genus of the tangent or normal bundle, 
and the $\hat{\cal L}({\cal R})$ the Hirzebruch polynomial. 
The physical gauge fields and curvatures are related to the
skew-hermitian ones in \bornina\ by rescaling with $-4i\pi^2\alpha'$. 
These expressions simplify drastically for sLag 3-cycles, where 
${\rm ch}({\cal F})\vert_{{\rm D}p}={\rm rk}({\cal F})$, the other 
characteristic classes become trivial and finally the only contribution in the
CS-term \bornina\ then comes from $C_{7}$. 

In the following we denote the homology class of Fix$(\o\sigma)$ by 
$\pi_{{\rm O}6} = [{\rm Fix}(\o\sigma)] \in H_3({\cal M}^{6})$ and 
the homology class of any given brane stack D$6_a$-brane by $\pi_a$. 
By our assumptions the $\pi_a$ are never invariant under $\o\sigma$
but mapped to image cycles $\pi'_a$. Therefore,
a stack of D6-branes is wrapped on that cycle by symmetry, too.
The RR charge cancellation can now easily be deduced by looking at the 
equation of motion of $C_{7}$
\eqn\equaasa{ {1\over \kappa^2}\,
d\star d C_{7}=\mu_6\sum_a N_a\, \delta(\pi_a)+
                    \mu_6\sum_a N_a\, \delta(\pi'_a)
        + \mu_6 Q_6\,  \delta(\pi_{{\rm O}6}),
}
where $\delta(\pi_a)$ denotes the Poincar\'e dual form of $\pi_a$, 
$\mu_p = 2\pi (4\pi^2\alpha')^{-(p+1)/2}$, and 
$2 \kappa^2= \mu_7^{-1}$. 
Upon integrating over ${\cal M}^6$ the RR-tadpole cancellation
condition becomes a relation in homology
\eqn\tadhom{
\sum_a  N_a\, (\pi_a + \pi'_a)-4\pi_{{\rm O}6}=0.
}
In principle it involves as many linear relations as there are
independent generators in $H_3({\cal M}^{6},\IR)$. But, of course,
the action of $\o\sigma$ on ${\cal M}^{6}$ also induces an action
$[\o\sigma]$ on the homology and cohomology. In particular,
$[\o\sigma]$ swaps $H^{2,1}$ and $H^{1,2}$, and 
the number of conditions is halved.

Similarly one can determine the disc level tension by 
integrating the Dirac-Born-Infeld effective
action. It is proportional to the
volume of the D-branes and the O-plane, so that the
disc level scalar potential reads
\eqn\susy{
{\cal V}=T_6\, {e^{-\phi_{4}}
\over \sqrt{{\rm Vol({\cal M}^{6})}}}
               \left( \sum_a  N_a \left( {\rm Vol}({\rm D}6_a) + 
      {\rm Vol}({\rm D}6'_a) \right) -4 {\rm Vol}({\rm O}6)\right). 
}
The potential is easily seen to be positive semidefinite and its 
vanishing imposes conditions on some of the moduli, freezing them 
to fixed values. Whenever the potential is non-vanishing, supersymmetry 
is broken and a classical vacuum energy generated by the net brane tension. 
It is easily demonstrated that the vanishing of \susy\ requires all 
the cycles wrapped by the D6-branes to be calibrated with respect to the same 
3-form as are the O6-planes. In a first step, just to conserve supersymmetry 
on their individual world volume theory, the cycles have to be 
calibrated at all, which leads to 
\eqn\dbi{
{\cal V}=T_6\, e^{-\phi_{4}} \left(
\sum_a{N_a \left| \int_{\pi_a} \widehat\Omega_3 \right|} +
 \sum_a{N_a \left| \int_{\pi'_a} \widehat\Omega_3 \right|} 
-4 \left| \int_{\pi_{{\rm O}6}} \widehat\Omega_3 \right |\right) .
}
Since $\widehat\Omega_3$ is closed, the integrals only depend on
the homology class of the world volumes of the branes and planes and 
thus the tensions also become topological. If we further demand that 
any single D$6_a$-brane conserves the same
supersymmetries as the orientifold plane the cycles 
must all be calibrated with respect to $\Re(\widehat\Omega_3)$. We can 
then write 
\eqn\dbic{
{\cal V}=T_6\,  e^{-\phi_{4}} 
\int_{\sum_a{N_a (\pi_a+\pi_a')
-4\pi_{{\rm O}6}}} \Re(\widehat\Omega_3) .
}
In this case, the RR charge and NSNS tension cancellation is 
equivalent, as expected in the supersymmetric situation .

\subsec{Massless closed string modes}

The action of $\Omega\o\sigma$ on the cohomology determines the 
spectrum of the closed string bulk modes as usual. 
One simply needs to consider all the massless 
fields of the ${\cal N}=2$ type IIA 
theory after compactification on the Calabi-Yau 3-fold and 
project out those that are odd under $\Omega\o\sigma$ when truncating  
to ${\cal N}=1$ supersymmetry. Before the projection there were
$h^{(1,1)}$ abelian vector multiplets and $h^{(2,1)}$
hypermultiplets. The $h^{(1,1)}$ vector multiplets
consist of one scalar coming from the  dimensional reduction of the
gravity field (the K\"ahler modulus), another scalar resulting
from the reduction of the NSNS 2-form and a four-dimensional
vector from the reduction of the RR 3-form along the 2-cycle.
If the $(1,1)$ form is invariant under
$\Omega\o\sigma$ an ${\cal N}=1$ chiral multiplet survives the
projection and if it is odd we get an ${\cal N}=1$
vector multiplet. Note, that the surviving chiral multiplets
still contain the complexified K\"ahler moduli. 
On the other hand the four scalars of the
$h^{(2,1)}$ hypermultiplets contain two scalars from the
ten-dimensional gravity field (the complex structure moduli)
equipped with two scalars arising from the dimensional reduction of
the RR 3-form along the two associated 3-cycles referring to
$H^{2,1}({\cal M}^3)$ and $H^{1,2}({\cal M}^3)$. Under the
$\Omega\o\sigma$ projection one of the two components of the complex
structure is divided out and moreover one linear combination of
the RR scalars survives, so that the former quaternionic
complex structure moduli
space gets reduced  to a complex moduli space of dimension $h^{(2,1)}$.

\subsec{Massless open string modes}

In this section we are going to present the most important input for
constructing intersecting brane world models of particle physics, the
formulae that determine the spectrum of the chiral fermions of the
effective theory in terms of topological data of the brane
configuration and the Calabi-Yau manifold.  Roughly speaking, at any
intersection point of two stacks of D6-branes a single chiral fermion
is localized \rangles, transforming in the bifundamental
representation of the two respective gauge groups. One needs to take
care that also non-chiral matter in sectors where intersection points
of opposite orientation cancel will generically become massive, so
that the massless spectrum is really determined by the intersection
numbers. As was mentioned already, the search for a viable model close
to the Standard Model particle content boils down to looking for
Calabi-Yau spaces with an involution $\o\sigma$ and an intersection
form for its 3-cycles that allows to realize the desired particle
spectrum at the intersections.

Catching up with the above discussion of the brane tension and the
induced scalar potential, we can say a bit more: If we want to
construct a supersymmetric intersecting brane world we need the
desired intersection pattern to be realized within a set of sLag
cycles all calibrated by the same 3-form, which makes the task a lot
harder. In the following we shall actually be presenting a model on
the Calabi-Yau defined by the Fermat quintic polynomial in $\IC\IP^4$
which is built out of calibrated cycles, but not all calibrated with
respect to the same calibration form.  In supersymmetric models, the
total massless spectrum is easily found by adding superpartners to the
fermions. Upon breaking supersymmetry, it is to be expected that all
fields except gauge bosons and chiral fermions will get masses through
interactions.

To obtain the chiral spectrum of a given set of D6-branes wrapped on cycles 
$\pi_a$ with their images on $\pi_a'$ and the O6-planes wrapped on 
$\pi_{{\rm O}6}$ a few considerations are necessary. 
The only novelty is that in addition to the standard operation by 
$\Omega$ a permutation of the branes and intersection points 
by $\o\sigma$ occurs, formally 
encoded in acting by a permutation matrix on the Chan-Paton labels 
that determine the representation under the gauge group.  
First note that the net 
number of self-intersections of any stack vanish, due to the 
anti-symmetry of the intersection form (denoted by $\circ$). 
Whenever a brane intersects its own image, there are two cases to distinguish: 
The intersection can itself be invariant under $\o\sigma$, such 
that the Chan-Paton labels are anti-symmetrized by $\Omega$. 
Alternatively, it can also be mapped to a second intersection, such that no 
projection applies and the symmetric and antisymmetric parts are kept. 
Finally, if any two different stacks intersect, there are always 
bifundamental representations localized at the intersection. 
According to these rules, the spectrum of
left-handed massless chiral fermions is shown in table 1. 
\vskip 0.8cm
\vbox{
\centerline{\vbox{
\hbox{\vbox{\offinterlineskip
\def\tablespace{height2pt&\omit&&
 \omit&\cr}
\def\tablerule{\tablespace\noalign{\hrule}\tablespace}

\hrule\halign{&\vrule#&\strut\hskip0.2cm\hfill #\hfill\hskip0.2cm\cr
& Representation  && Multiplicity &\cr
\tablerule
& $[{\bf A_a}]_{L}$  && ${1\over 2}\left(\pi'_a\circ \pi_a+\pi_{{\rm O}6} \circ \pi_a\right)$   &\cr
\tablerule
& $[{\bf S_a}]_{L}$
     && ${1\over 2}\left(\pi'_a\circ \pi_a-\pi_{{\rm O}6} \circ \pi_a\right)$   &\cr
\tablerule
& $[{\bf (\o N_a,N_b)}]_{L}$  && $\pi_a\circ \pi_{b}$   &\cr
\tablerule
& $[{\bf (N_a, N_b)}]_{L}$
&& $\pi'_a\circ \pi_{b}$   &\cr
}\hrule}}}}
\centerline{
\hbox{{\bf Table 1:}{\it ~~ Chiral fermion  spectrum in $d=4$}}}
}
\vskip 0.5cm
\noindent
The above classification can be obtained directly from string 
amplitudes when a CFT 
description is available, e.g. in the orbifold limit, while at large
volume one can apply the Atiyah-Singer index theorem to infer the 
zero-modes of the Dirac operator. This is actually a tautology, since 
the number of chiral modes is given as an integral over the point-like 
common world volume of any pair of D6-branes with a trivial integrand, 
\eqn\atiyah{
\int_{{\rm D}6_a \cap {\rm D}6_b}
 {{\rm ch}({\cal F}_a) \wedge {\rm ch}({\cal F}^*_b) \wedge \hat A({\cal R})} = 
{\rm rk}({\cal F}_a){\rm rk}({\cal F}_b) 
\int_{{\cal M}^6} \delta(\pi_a) \wedge \delta(\pi_b) , 
}
which only counts the intersection numbers again. 
In the mirror symmetric type IIB picture chirality is in fact induced exclusively 
by the non-triviality of the gauge and spin connection. 

Due to the topological nature of the chiral spectrum table 1 should
hold for every smooth Calabi-Yau manifolds and even the
six-dimensional torus \rbgkl.  Little can be said about the fate of
the D-brane setting away from the limit of classical geometry, when
venturing into the interior of the K\"ahler moduli space, where
potentials may be generated. Therefore, the configuration will in
general not be stable, but the important point is, whenever the
setting is describable purely in terms of D6-branes on sLag 3-cycles
table 1 applies.
 
To make a first check of the consistency of the spectrum, 
the non-abelian gauge anomaly $SU(N_a)^3$ 
\eqn\anapp{
A_{\rm non-abelian} \sim \pi_a\circ \pi_a
}
vanishes due to the antisymmetry of the intersection form. 

\subsec{The Quintic}

Now that we have collected the machinery to construct intersecting 
brane worlds on general Calabi-Yau 3-folds, we proceed to discuss 
the example of the quintic. It is probably the most studied and best 
understood example of a hypersurface in a projective space. It even 
appears that there are no other examples known in the literature 
where a sufficiently large class of sLag cycles has been found and 
their intersection form classified. 

One defines the Fermat quintic by the following hypersurface in $\IC\IP^4$
\eqn\fermatquintic{
{\cal Q}:\  \sum_{i=1}^{5} z_i^5 = 0
  \quad \subset \IC \IP^4 .
}
It has the obvious involution from the complex conjugation of 
the coordinates $z_i\to \o{z}_i$ as a symmetry. 
The fixed points of $\o\sigma$ are the real 
quintic $\sum_{i=1}^{5} x_i^5 = 0 \subset \IR\IP^4$, topologically
a sLag $\IR\IP^3$.
As a further symmetry of the Fermat quintic a $\ZZ_5^5$ acts via
\eqn\quinticaction{
  z_i \mapsto \omega^{k_i} z_i
  \qquad \omega=e^{2\pi i\over 5},~k_i\in \ZZ_5 .
}
We can use this symmetry to generate a whole class of sLag cycles from 
the one prototype above. Because the diagonal $\ZZ_5$ is trivial, 
this produces $5^4=625$ different minimal $\IR\IP^3$, labeled 
by the integers $k_i$ and defined by 
\eqn\quinticRPs{
  \left|k_2,k_3,k_4,k_5\right>
  \mathrel{\buildrel{\rm def}\over=}
  \Big\{
  [x_1:\omega^{k_2}x_2:\omega^{k_3}x_3:\omega^{k_4}x_4:\omega^{k_5}x_5]
  \Big| x_i\in \IR,~\sum_{i=1}^5 x_i^5=0
  \Big\}
}
The only information further needed is their intersection form, 
determined in \rbdlr. The intersection of any cycle 
$\left|k_2,k_3,k_4,k_5\right>$ with the 
one $\left|1,1,1,1\right>$ is given by the 
coefficient of the monomial
$g_2^{k_2}g_3^{k_3}g_4^{k_4}g_5^{k_5}$ in
\eqn\quinticIntersect{
  I_{\IR\IP^3} =
  \prod_{i=1}^5 \big( g_i + g_i^2 - g_i^3 - g_i^4 \big)
  ~\in
  \ZZ[g_1,g_2,g_3,g_4,g_5]
  \Big/
  \big< g_i^5=1, \prod g_i=1 \big> . 
}
All the other intersection numbers are then obtained by applying the
$\ZZ_5^4$ symmetry. The ensuing intersection matrix
$M\in {\rm Mat}(625,\ZZ)$ has rank $204=b_3$, so the
$\left|k_2,k_3,k_4,k_5\right>$ generate
the full $H_3({\cal Q};\IR)$.

Of course, only one fifth of the minimal $\IR\IP^3$ are $\Re(\Omega_3)$ 
calibrated, while the others are calibrated with respect to 
$\Re (\omega^k \Omega_3)$. To determine the
$\Re(\Omega_3)$ sLags we need to know how $\ZZ_5^4$ acts on the
holomorphic volume form. From the residue formula
\eqn\residueformula{
  2 \pi i \Omega_3 =
  \oint_\Gamma
  { {
    \epsilon^{i_1 \cdots i_5} z_{i_1}
    {\rm d}z_{i_2}\wedge {\rm d}z_{i_3}\wedge
    {\rm d}z_{i_4}\wedge {\rm d}z_{i_5}
  } \over {
    \sum_{i=1}^5 z_i^5
  } }
}
it is evident that $\Omega_3$ transforms as 
$\Omega_3 \mapsto (\prod_i\omega^{k_i}) \Omega_3$,
such that the
$\left|k_2,k_3,k_4,k_5\right>$ are calibrated with respect to $\Re(\Omega_3)$ 
precisely if $\sum_{i=2}^5 k_i = 0~{\rm mod}~5$.
Using the intersection matrix for these 125 sLags one can check that 
they generate a $101$-dimensional subspace of $H_3({\cal Q})$. 
As was discussed at length, in order to construct any supersymmetric 
brane world model, it would be necessary to use D6-branes wrapping these 
125 cycles only. Only in this case the scalar potential generated by the 
tension of the branes would be balanced by the negative tension of the 
O6-planes. Unfortunately, this turns out to be impossible with the present 
class of sLags, since it is found that the 
intersections among themselves all vanish. Therefore, a chiral spectrum cannot 
be reconciled with a supersymmetric groundstate. The validity of this 
statement is in fact rather limited. One cannot even conclude that 
a supersymmetric brane world model is not accessible on the Fermat quintic, 
because it may well happen to exist within another set of sLag cycles. 
Not to mention that the general quintic may have points in its moduli 
space where another involution can be used to define $\o\sigma$ and completely 
different sets of sLags exist.

\subsec{The Quintic Standard Model}

We have seen that using only the 3-cycles in \quinticRPs\ 
we cannot obtain interesting brane configurations
if we want to preserve supersymmetry, which would restrict us to only 
using the 3-cycles in \quinticRPs. However, it is indeed possible to 
construct a model with the correct intersection numbers by 
dropping the requirement of supersymmetry, as we shall demonstrate in the 
following. The breaking of supersymmetry is still of a special and 
somehow weak nature, since the individual stacks still respect  
some supersymmetry generators, just not all the same. This can have  
interesting consequences for the dynamics of the effective gauge theory.

A further generalization of the set of sLags is needed since 
all intersection numbers of the $625$ minimal
$\IR\IP^3$ are in the range $-2,\,\dots\, ,2$, while we need $\pm 3$
for some cycles to reproduce the three generation structure 
of the Standard Model fermion spectrum. 
So we must use linear combinations. 
There are some subtleties with this step. First notice that 
by adding the sLag cyles in homology, their volumes also just 
add up in accordance with the topological nature of their tension. 
Furthermore, if the two cycles in question are calibrated with respect to the 
same calibration 3-form, their sum in homology, represented by the 
geometrical union of the two submanifolds, will as well be calibrated by this 3-form. 
But the union of two sLag submanifolds will usually not be a smooth and 
connected submanifold itself, and thus cannot simply be wrapped by any 
single stack of D-branes. 

Let us illustrate this problem with a simple 
example of sLag submanifolds on a four-dimensional torus 
$T^4=T_1^2 \times T_2^2$. Take the two $T_I^2$ to be given by square tori of 
volume 1. Now consider 2-cycles calibrated by 
$\Re(\Omega_2)=dx_1\wedge dx_2-dy_1\wedge dy_2$, given by 
lines on any one of the two $T_I^2$ satisfying $\varphi_1+\varphi_2=0$ 
for their relative angles $\varphi_I$ with respect to the $x_I$-axis 
on each $T_I^2$. Any such line is 
defined homologically by specifying the 
two 1-cycles it wraps on the $T_I^2$. So let us denote them by two times two 
integers $(n_a^I,m_a^I)$, $a$ labeling the stacks. 
The calibration condition is solved by $n_a^1=n_a^2,\, m_a^1=-m_a^2$ and 
the volume of any such sLag is given by Vol$_a=(n_a^1)^2+(m_a^1)^2$. 
One may pick a basis in homology by using all 2-cycles with two entries 
0 and 1 respectively.  
Now compare for instance the cycles $(1,0;1,0)+(0,1;0,-1)$ and 
$(1,1;1,-1)$, both with total volume 2. The first is the sum of two sLag 
cycles of volume 1, each projecting onto either the $x_I$-axes or 
the $y_I$-axes of the $T_I^2$, 
while the second one is the product of the diagonals. 
This relation closely resembles what we are seeking in the Calabi-Yau case 
of the quintic. We add up two sLags of equal volume 1. The union 
of the two submanifolds first consists of two components intersecting 
in a point at the origin of the two $T^2$. But we also find a smooth 
sLag in the homology class $(1,1;1,-1)$ which is the sum of the two 
individual cycles plus a component that is not calibrated by $\Re(\Omega_2)$. 
This demonstrates that a calibrated cycle cannot be decomposed into 
basis elements all calibrated by the same calibration form. Neither 
can one expect sums of of calibrated basis cycles to be represented by 
smooth and connected sLag submanifolds. 

But there is a physical argument for the existence of a smooth connected 
representative in exactly the class of the sum of two sLags, that 
may apply, whenever the two original cycles intersect. 
At the intersection point there will be localized 
scalar moduli in the bifundamental of the gauge groups 
on the two stacks. Turning on vacuum expectation values for these 
corresponds geometrically to deforming the singular geometry at 
the intersection into a smooth submanifold, where the two 
components have joined together. The product gauge group is actually broken 
to the diagonal and the decomposition of the bifundamental contains 
a neutral scalar which parameterizes the deformation. 
We therefore expect a smooth connected cycle of minimal volume to exist in 
the homology class of the sum of any two intersecting sLag representatives. 
As in the toroidal example above, it may not fall into the original class of 
sLag cycles, and there is no control over its calibration condition. On the 
other hand, cycles which do not intersect in homology may be disentangled 
geometrically by smooth deformations. D-branes wrapped on such 
classes have to be expected to decay into disjoint components, i.e. multiple 
stacks of D-branes. 
We shall now use this criterion to improve the example for a Standard Model 
brane world given in \BlumenhagenWN. 

In order to realize the three generation Standard Model spectrum on an 
intersecting brane world on the Fermat quintic we then employ a 
combination of the sLag $\IR\IP^3$ cycles and linear combinations of such 
cycles which intersect each other. Any single stack is then still expected to 
maintain four linearly realized supersymmetry generators, 
but not all of them the same. Concretely, we have an 
O6-plane on the cycle $\pi_{{\rm O}6}=\left|0,0,0,0\right>$ and can choose to 
wrap D6-branes on the following 3-cycles
%
\eqn\quinticcycles{\eqalign{
  \eqalign{
    \pi_a =&~    \pi_c - \pi_d - \left|0,2,1,4\right> - \left|0,3,4,1\right> \cr
    \pi_b =&~    \left|0,3,1,1\right> \cr
    \pi_c =&~    \left|1,4,3,4\right> + \left|4,4,3,2\right> \cr
    \pi_d =&~    \left|0,3,0,3\right> - \left|2,0,3,4\right> \cr
  } \qquad \Rightarrow \qquad
  \eqalign{
    \pi_a'  =&~  \pi_c' - \pi_d' - \left|0,3,4,1\right> - \left|0,2,1,4\right> \cr
    \pi_b'  =&~    \left|0,2,4,4\right> \cr
    \pi_c'  =&~    \left|4,1,2,1\right>+ \left|1,1,2,3\right> \cr
    \pi_d'  =&~    \left|0,2,0,2\right>- \left|3,0,2,1\right> . \cr
  }
}}
On the whole we have been able to find a fairly large number of hundreds 
of configurations with the same 
chiral fermion spectrum meeting the requirements. 
The one given above looks rather complicated because it is designed 
to meet the additional condition 
\eqn\hypercond{
(\pi_a-\pi_a') -(\pi_c-\pi_c')+(\pi_d-\pi_d') = 0 
}  
for a massless hypercharge gauge boson, 
as to be explained in section 3.3. 
It is also straightforward to check that the homology classes are primitive, 
i.e. not a multiple of some other class in $H_3({\cal Q},\ZZ)$, 
by finding at least one other cycle such that the 
intersection number with this other one is $\pm 1$.
This was an important consistency requirement for toroidal models. 
The intersection numbers of the given $3$-cycles are
shown in table 2.
The matrix has the following symmetries 
\eqn\intersectsymmetries{\eqalign{
  \pi_i \circ \pi_j = -\pi_j \circ \pi_i = \pi_j' \circ \pi_i' = -\pi_i' \circ \pi_j' , \cr
  \pi_i \circ \pi_j' = \pi_j \circ \pi_i' = -\pi_i' \circ \pi_j = \pi_j \circ
  \pi_i'.
}}
Table 2 just reproduces the ``intersection numbers of the 
Standard Model'' as proposed in \rimr: If one wraps 
$3$ branes on $\pi_a$, $2$ branes on $\pi_b$, and a
single brane on $\pi_c$ and $\pi_d$, the gauge group
is $U(3)\times U(2)\times U(1)^2$ before performing any anomaly analysis. 
\vskip 0.8cm
\vbox{
\centerline{\vbox{
\hbox{\vbox{\offinterlineskip
\def\tablealign{&\omit&&\omit&&\omit&&\omit&&\omit&&
     \omit&&\omit&&\omit&&\omit&&\omit&\cr}
\def\tablespace{height2pt\tablealign}
\def\tablerule{\tablespace\noalign{\hrule}\tablespace}

\def\tableruleB{\tablespace\noalign{\hrule}height0.5mm\tablealign\noalign{\hrule}}
\hrule\halign{
&\vrule#&\strut\hskip0.2cm\hfill $#$\hfill\hskip0.2cm
&\vrule#\hskip0.5mm\vrule&\strut\hskip0.2cm\hfill $#$\hfill\hskip0.2cm
&\vrule#&\strut\hskip0.2cm\hfill $#$\hfill\hskip0.2cm
&\vrule#&\strut\hskip0.2cm\hfill $#$\hfill\hskip0.2cm
&\vrule#&\strut\hskip0.2cm\hfill $#$\hfill\hskip0.2cm
&\vrule#&\strut\hskip0.2cm\hfill $#$\hfill\hskip0.2cm
&\vrule#&\strut\hskip0.2cm\hfill $#$\hfill\hskip0.2cm
&\vrule#&\strut\hskip0.2cm\hfill $#$\hfill\hskip0.2cm
&\vrule#&\strut\hskip0.2cm\hfill $#$\hfill\hskip0.2cm
&\vrule#&\strut\hskip0.2cm\hfill $#$\hfill\hskip0.2cm
\cr
& \circ
&&  \pi_a &&  \pi_b &&  \pi_c &&  \pi_d
&& \pi_a' && \pi_b' && \pi_c' && \pi_d'
&& \pi_{{\rm O}6}
&\cr
\tableruleB
& \pi_a  &&  0 &&  -1 && 3 &&  0 &&  0 &&  -2 && 3 &&  0 &&  0 &\cr
\tablerule
& \pi_b  && 1 &&  0 &&  0 &&  0 &&  -2 &&  0 &&  0 && 3 &&  0 &\cr
\tablerule
& \pi_c  &&  -3 &&  0 &&  0 && 3 && 3 &&  0 &&  0 &&  -3 &&  0 &\cr
\tablerule
& \pi_d  &&  0 &&  0 &&  -3 &&  0 &&  0 && 3 && -3 &&  0 &&  0 &\cr
\tablerule
& \pi_a' &&  0 && 2 &&  -3 &&  0 &&  0 && 1 &&  -3 &&  0 &&  0 &\cr
\tablerule
& \pi_b' && 2 &&  0 &&  0 &&  -3 &&  -1 &&  0 &&  0 &&  0 &&  0 &\cr
\tablerule
& \pi_c' &&  -3 &&  0 &&  0 && 3 && 3 &&  0 &&  0 &&  -3 &&  0 &\cr
\tablerule
& \pi_d' &&  0 &&  -3 && 3 &&  0 &&  0 &&  0 && 3 &&  0 &&  0 &\cr
\tablerule
&\pi_{{\rm O}6}&&  0 &&  0 &&  0 &&  0 &&  0 &&  0 &&  0 &&  0 &&  0 &\cr
}\hrule}}}}
\centerline{
\hbox{{\bf Table 2:}{\it ~~ Intersection numbers}}}
}
\vskip 0.5cm
\noindent
The intersection matrix then produces the bifundamental fermions of 
the Standard Model and nothing else, as shown in table 3. Since $\pi_i\circ
\pi_i'=\pi_i \circ \pi_{{\rm O}6}=0$ there are no chiral fermions
in the symmetric or antisymmetric representations of the gauge
groups. 
\vskip 0.8cm
\vbox{
\centerline{\vbox{
\hbox{\vbox{\offinterlineskip
\def\tablespace{height2pt&\omit&&\omit&&\omit&&\omit&&
 \omit&\cr}
\def\tablerule{\tablespace\noalign{\hrule}\tablespace}

\hrule\halign{&\vrule#&\strut\hskip0.2cm\hfil#\hfill\hskip0.2cm\cr
\tablespace
& Sector && Field &&  $SU(3)\times SU(2)\times U(1)^4$ && $U(1)_Y$
&& Multiplicity &\cr
\tablerule
& $(ab)$ && $Q_L$ && $({\bf 3},{\bf 2})_{(1,-1,0,0)}$ &&  $1/3$
&& $1$ & \cr
\tablespace
& $(a'b)$ && $Q_L$ &&  $({\bf 3},{\bf 2})_{(1,1,0,0)}$ &&  $1/3$
&& $2$ & \cr
\tablespace
& $(ac)$ && $U_R$ && $(\o{\bf 3},{\bf 1})_{(-1,0,1,0)}$ &&
$-4/3$ && $3$ & \cr
\tablespace
& $(a'c)$ && $D_L$ && $(\o{\bf 3},{\bf 1})_{(-1,0,-1,0)}$ &&
$2/3$ && $3$ & \cr
\tablerule
& $(b'd)$ && $L_L$ && $({\bf 1},{\bf 2})_{(0,-1,0,-1)}$ &&  $-1$
&& $3$ & \cr
\tablespace
& $(cd)$ && $E_R$ && $({\bf 1},{\bf 1})_{(0,0,-1,1)}$ &&  $2$ &&
$3$ & \cr
\tablespace
& $(c'd)$ && $N_L$ && $({\bf 1},{\bf 1})_{(0,0, 1,1)}$ &&  $0$ &&
$3$ & \cr
\tablespace}\hrule}}}}
\centerline{
\hbox{{\bf Table 3:}{\it ~~ Chiral left-handed fermions for the 3 generation
model.}}}
}
\vskip 0.5cm
\noindent
The fermion spectrum leaves two of the abelian factors free of anomalies. One is 
the Standard Model hypercharge and given by
\eqn\hyper{  U(1)_Y={1\over 3} U(1)_a - U(1)_c + U(1)_d .
}
In addition, the quantum number $B-L$ is gauged as well, since the spectrum of the 
Standard Model with right-handed neutrinos leaves it anomaly-free. 
The other two of the 
four abelian factors are anomalous and decouple through a generalized Green-Schwarz 
mechanism, leaving the Standard Model gauge group with one extra $U(1)$. 
In order to ensure that the second gauge boson does get a mass, 
while the hypercharge gauge boson really remains massless,  
one additionally has to analyze the couplings to the various axions that descend 
from the dimensional reduction of the RR 5-form potential. 
We shall come to this point later and show that the extra restriction 
can be satisfied. 
  
An important point to notice is that the model so far does have a 
non-vanishing RR tadpole. The sum of the homology classes
\eqn\quintictadhom{\eqalign{
3 (\pi_a+\pi_a') + 2(\pi_b+\pi_b') + (\pi_c+\pi_c') + (\pi_d+\pi_d')
-4 \pi_{{\rm O}6}
}}
does not vanish. To cancel the RR charge without changing the spectrum, 
one can easily introduce a hidden brane sector that
carries the right charge to cancel the tadpole but does not intersect the
Standard Model branes, so there is no chiral matter charged under both
the visible and the hidden gauge group. 

\subsec{Large transverse volume}

In order to reconcile the string scale 
supersymmetry breaking which occurs in the model just 
presented with the hierarchy problem one may refer to a large extra dimension 
scenario with a fundamental string scale of the order of a TeV \refs{\radd,\raadd}. 
The four-dimensional gauge couplings $g_a$ and Planck mass $M_{pl}$ are obtained by 
dimensional reduction from the fundamental string scale $M_s$ and string 
coupling $g_s$ according to 
\eqn\copl{
{1\over g_a^2} \sim {1\over g_s} {{\rm Vol}({\rm D}6_a) \over l_s^3} , \quad 
M_{pl}^2 \sim {M_s^2 \over g_s^2} {{\rm Vol}({\cal M}^6) \over l_s^6} . 
}
The conditions for a high effective Planck scale with a TeV string scale 
can then be phrased 
\eqn\tevscale{
{{\rm Vol}({\cal M}^6) \over l_s^6} \gg  {{\rm Vol}({\rm D}6_a)^2 \over l_s^6}. 
}
Numerical values for plausible assumptions have been discussed in the
introduction already.

\newsec{Scalar potential, anomalies and gauge boson masses}

Non-supersymmetric brane configurations are in general unstable.
On the one hand, depending on the intersection angles there can be tachyons
localized at the intersection points. Phenomenologically it was
suggested that these tachyons might be interpreted as Standard
Model Higgs fields \refs{\rbachas,\rbgklnon}, where
in particular in \rqsusyb\  it was demonstrated that the
gauge symmetry breaking is consistent with this point of view.
In general we expect that tachyons can well be avoided in 
a large subset of the parameter space, as was due in the 
simpler setting of toroidal compactifications. 
On the other hand, even if tachyons are absent one generally
faces uncanceled NSNS tadpoles, which might destabilize the
configuration \refs{\berlin,\rott,\BlumenhagenUA,\belrab}.
In \berlin\ it was shown that for appropriate
choices of the D-branes the complex structure moduli
can be stabilized by the induced tree level potential.
The stabilization of the dilaton remains a major challenge
as in all non-supersymmetric string models.

For supersymmetric intersecting brane worlds we can
expect much better stability properties.
First tachyons are absent in these models due to the
Bose-Fermi degeneracy. However, since for orientifolds on Calabi-Yau spaces
the configuration only preserves ${\cal N}=1$ supersymmetry,
in general non-trivial F-term and D-term potentials
can be generated. 

\subsec{F-term superpotential and D-terms}

There are strong restrictions known for the contributions that can 
give rise to corrections to the effective ${\cal N}=1$ superpotential 
of a type II compactification on a Calabi-Yau 3-fold with D6-branes and
O6-planes on supersymmetric 3-cycles. 
The standard arguments about the non-renormalization of the superpotential by 
string loops and world sheet $\alpha'$ corrections apply. 
The only effects then 
left are non-perturbative world sheet corrections, open and closed 
world-sheet instantons. In general, 
these are related to non-trivial $\IC\IP^1$ and 
$\IR\IP^2$ with boundary on the O6-plane in the Calabi-Yau manifold 
for the closed strings 
and discs with boundary on the D6-branes for open strings.
In fact, only the latter contribute to the superpotential. 
The typical form for the superpotential thus generated is known, but 
explicit calculations are only available for non-compact models. Usually, they 
make use of open string mirror symmetry arguments. 
In many cases, there is an indication that the non-perturbative 
contributions to the superpotentials 
tend to destabilize the vacuum, and it would be a tempting 
task to determine a class of stable ${\cal N}=1$ supersymmetric
intersecting brane models.

The tension of the D6-branes and O6-planes in addition introduces a 
vacuum energy 
which is described in terms of D-terms in the language of ${\cal N}=1$ 
supersymmetric field theory. These depend only on the complex structure 
moduli and do not affect the K\"ahler parameter of the background. 
The most general form for such a potential is given by
\eqn\dterm{
{\cal V}_{\rm D-term} =
\sum_a {1\over 2g_a^2}\biggl(\sum_i q^i_a|\phi_i|^2+\xi_a\biggr)^2\, ,
}
with $g_a$ the gauge coupling of a $U(1)_a$, 
$\xi_a$ the FI parameter, and the
scalar fields $\phi_i$ are the superpartners of some bifundamental fermions at the 
intersections. They become massive or tachyonic 
for non-vanishing $\xi_a$, depending on their charges $q^i_a$. 
Due to the positive definiteness of the D-term, 
${\cal N}=1$ supersymmetry will only be unbroken in the vacuum,
if the potential vanishes. 

This requirement can be compared to the calibration condition for 
the 3-cycles wrapped by the branes. If they are just individually sLag, one can use 
\dbic\ to write  
\eqn\dbia{
{\cal V}=
2\,T_6\, e^{-\phi_4}\,
\sum_a N_a \left( \left| \int_{\pi_a} \widehat\Omega_3 \right|-
              \int_{\pi_a} \Re(\widehat\Omega_3) \right)   .
}
To apply \dterm\ we have to use the properly normalized gauge coupling 
\eqn\gaugec{
{1\over g_{U(1)_a}^2}={N_a\over g_a^2}={N_a\, M_s^3\over (2\pi)^4}\,
                     e^{-\phi_{4}}
                \left| \int_{\pi_a} \widehat\Omega_3 \right|.
                }
Hence, the FI-parameter $\xi_a$ can be identified as
\eqn\fiterm{
\xi^2_a={M_s^4 \over 2\pi^2} { \left| \int_{\pi_a} \widehat\Omega_3 \right|-
               \int_{\pi_a} \Re(\widehat\Omega_3) \over
                \left| \int_{\pi_a} \widehat\Omega_3 \right|} ,
}
which vanishes precisely if the overall tension of the branes and planes cancels out, 
i.e. if all are calibrated with respect to the same 3-form. 
Since the FI-term is not a holomorphic quantity one expects
higher loop corrections to the classical result \fiterm.

\subsec{Anomalies}

While the non-abelian anomalies of the chiral fermion spectrum given in 
table 3 vanish in any case, the mixed and abelian anomalies do not. 
Their cancellation requires various axions to participate in a generalized 
Green-Schwarz mechanism to render the theory consistent. One can actually 
check in detail that the relevant couplings match with the 
anomalous contribution 
\eqn\mixeda{
A_{ab}={N_a \over 2} ( - \pi_a + \pi_a' ) \circ \pi_b 
}
from the chiral matter spectrum. It is in fact sufficient to consider the 
case of the $U(1)_a-SU(N_b)^2$ diagrams for $a\not=b$. 

To do so, it is useful to define an integral basis
for $H_3({\cal M}^6,{\ZZ})$, given by 3-cycles $\alpha^I,\, \beta_J$, 
$I,J=0,\dots ,h^{(2,1)}$ with the property
$\alpha^I\circ\alpha^J=\beta_I\circ\beta_J=0$ and
$\alpha^I\circ\beta_J=\delta^I_J$.
We then expand the $\pi_a$ in terms of the
basis cycles $\alpha^I$ and $\beta_J$,
\eqn\expand{
\pi_a=e^a_I\alpha^I+m_a^J\beta_J\, ,
}
with integers $e^a_I$, $m_a^J$, and similarly for $\pi_a'$. 
The general Chern-Simons couplings reduce to 
\eqn\cscoupl{
\int_{\IR^{1,3}\times(\pi_a+\pi_a')}  C_3\wedge {\rm Tr}\left(F_a\wedge F_a\right) , \quad\quad
\int_{\IR^{1,3}\times(\pi_a+\pi_a')}  C_5\wedge {\rm Tr}\left(F_a\right) . 
}
The four-dimensional axions $\Phi_I$ and the dual 
2-form $B^I$, $I=0,\, ...\, , h^{(2,1)}$ are 
\eqn\formsl{\eqalign{ \Phi_I=\int_{\alpha^I} C_3, \quad\quad
                      \Phi^{I+h^{(2,1)}+1}=\int_{\beta_I} C_3, \cr
                      B^I=\int_{\beta_I} C_5, \quad\quad
                      B_{I+h^{(2,1)}+1}=\int_{\alpha^I} C_5 .}
}
More precisely, $(d\Phi_I, dB^I)$ and
$(d\Phi^{I+h^{(2,1)}+1},dB_{I+h^{(2,1)}+1})$ are Hodge dual to each other 
in four dimensions.
The general couplings \cscoupl\ can now be expanded 
\eqn\dddd{\eqalign{
\int_{\IR^{1,3}\times(\pi_a+\pi_a')}  C_3\wedge
       {\rm Tr}\left(F_a\wedge F_a\right)=&
\sum_I \left( e_a^I+(e_a^I)'\right) \int_{\IR^{1,3}} \Phi_I \wedge {\rm Tr}\left(F_a\wedge
                 F_a\right) \cr
&+\sum_I \left( m^a_I+(m^a_I)'\right) \int_{\IR^{1,3}}
\Phi^{I+h^{(2,1)}+1}
    \wedge {\rm Tr}\left(F_a\wedge F_a\right), \cr
\int_{\IR^{1,3}\times(\pi_a+\pi_a')}  C_5\wedge {\rm Tr}\left(F_a\right)
     =&N_a \sum_I \left( m^a_I-(m^a_I)'\right) \int_{\IR^{1,3}} B^I \wedge F_a \cr
&+N_a \sum_I \left( e_a^I-(e_a^I)'\right) \int_{\IR^{1,3}}
B_{I+h^{(2,1)}+1} \wedge F_a . }
 }
Adding up all terms for the  $U(1)_a-SU(N_b)^2$
anomaly, one finds 
\eqn\aadito{\eqalign{
A^{(2)}_{ab} ~\sim&~ N_a \sum_I \left( 
\left(e_a^I+(e_a^I)'\right)
               \left(m^b_I-(m^b_I)'\right)+
\left(m^a_I+(m^a_I)'\right)\left(e_b^I-(e_b^I)'\right) \right)
~\sim \cr
\sim&~ 2N_a \left( \pi_a - \pi'_a\right)\circ\pi_b, 
}}
which has just the right form to cancel the anomalous contribution \mixeda\
of the chiral fermions.

\subsec{Gauge boson masses}

The starting gauge group of our model contained four abelian factors, 
of which two anomalous ones get massive by the Green-Schwarz mechanism. 
A very important input to decouple some of the superfluous 
abelian factors from the 
unbroken and anomaly-free gauge group is the occurance of St\"uckelberg mass
terms from axionic couplings even without an anomaly \rimr. 
These couplings arise from the reduction of the RR 5-form in \dddd. 
In principle, $C_5$ can be reduced along any 3-cycle to produce $2h^{(2,1)}+2$
2-forms, as stated above. 
Again, only two of them are effectively 
involved in the Green-Schwarz mechanism to give masses 
to the gauge bosons of the two anomalous $U(1)$. 
At first sight, it then looks unlikely that any of the other two gauge bosons 
could evade getting a mass, being outnumbered by axions. 
One can rewrite the coupling term for the 2-forms 
$B^I$ in the Lagrangian like
\eqn\sumcoupl{
M_I^a B^I \wedge F_a = \sum_I B^I \wedge \sum_a N_a (m^a_I - (m^a_I)') F_a 
}
with a $4\times (h^{(2,1)}+1)$ matrix $M_I^a$, and similarly a second such 
term for the $B_{I+h^{(2,1)}+1}$. 
The requirement to have a massless 
hypercharge gauge boson now translates into 
\eqn\condition{
M_I^a (1,0,-3,3)_a= (0,\, ...\, ,0)^I .
} 
This implies that the $\beta_I$-components of the cycles 
$N_a(\pi_a-\pi_a')$ are 
required to be linearly dependent. A similar relation has to hold for 
the couplings to the 2-forms $B_{I+h^{(2,1)}+1}$ and the 
$\alpha^I$-components, as well. This is very suggestive: Every independent 
cycle introduces one axionic coupling and the axion is eaten by one of the 
$U(1)$ gauge bosons. In order to have a surviving gauge boson, a linear 
relation needs to hold between the cycles. If the second $U(1)$ is meant to 
decouple \condition\ should be the only linear relation among 
the $N_a(\pi_a-\pi_a')$.   
For the toroidal case one can check that 
the conditions derived in \rimr\ in the dual type IIB picture 
precisely reproduce \condition. 
The concrete model defined in \quinticcycles\ does in fact satisfy \condition\ or  
equivalently \hypercond\ and in addition $M_I^a$ has rank equal to 3. 
This means that the only unbroken abelian gauge symmetry is the 
hypercharge $U(1)_Y$. 

\newsec{Intersecting brane worlds in six dimensions}

The methods developed above for constructing four-dimensional intersecting 
brane world models on smooth Calabi-Yau backgrounds can also be applied 
to orbifolds. In this case one first needs to resolve the singular geometry 
in order to be able to compare to the classical data encoded in the 
intersection numbers. In this section we demonstrate the elegance and 
technical simplicity of the construction for six-dimensional K3-orbifolds, 
which is slightly simpler than Calabi-Yau-orbifolds, and 
via the six-dimensional constraints on 
anomaly cancellation offers an excellent check on the consistency of 
the results. Of course, some modifications need to be applied to the 
four-dimensional prescriptions in order to adapt to the K3. 
We are not going to explain everything in detail, 
but refer the reader to \BlumenhagenWN\ for more instructions and 
proper definitions.    
Further extensions to F-theory and M-theory vacua were also  
discussed in this reference. 

\subsec{K3 compactification}

In general, the compactification of type IIB on 
a K3 leaves ${\cal N}=(0,1)$ supersymmetry in six dimensions. 
The closed string fields usually do not cancel the gravitational anomaly 
of the graviton multiplet by their own. The well known condition 
\eqn\anomgrav{
n_{\rm H}-n_{\rm V}+29\,  n_{\rm T}=273 
}
determines whether the irreducible $R^4$ coefficient cancels out, 
when taking the open string states into account as well. 

The involution $\o\sigma$ now leaves fixed sLag 2-cycles in the K3, 
which are wrapped by O7-planes, whose charge is then canceled by D7-branes, 
according to the cancellation condition 
\eqn\tadpole{
\sum_a  N_a\, (\pi_a +\pi'_a)-
          8\,   \pi_{{\rm O}7}=0.
}
The gauge group supported by the various stacks is again given by 
\gauge, while the chiral spectrum can be determined in complete analogy 
to the four-dimensional case. It is summarized in table 4, 
the subscripts denoting the
representation under the little group $SO(4)\simeq SU(2)\times
SU(2)$, which is to be flipped for a negative intersection number.
\vskip 0.8cm
\vbox{
\centerline{\vbox{
\hbox{\vbox{\offinterlineskip
\def\tablespace{height2pt&\omit&&
 \omit&\cr}
\def\tablerule{\tablespace\noalign{\hrule}\tablespace}

\hrule\halign{&\vrule#&\strut\hskip0.2cm\hfill #\hfill\hskip0.2cm\cr
& Representation  && Multiplicity &\cr
\tablerule
& [{\bf Adj}]$_{(1,2)}$  && $\pi_a\circ \pi_a$   &\cr
\tablerule
& $[{\bf A_a+\o{A}_a}]_{(1,2)}$  && ${1\over 2}\left(\pi_a\circ
\pi'_a+\pi_a\circ \pi_{{\rm O}7}\right)$   &\cr
\tablerule
& $[{\bf S_a+\o{S}_a}]_{(1,2)}$
     && ${1\over 2}\left(\pi_a\circ \pi'_a-\pi_a\circ \pi_{{\rm O}7}\right)$   &\cr
\tablerule
& $[{\bf (N_a,N_b)+(\o N_a,\o N_b)}]_{(1,2)}$  && $\pi_a\circ \pi_{b}$   &\cr
\tablerule
& $[{\bf (N_a,\o N_b)+( \o N_a, N_b)}]_{(1,2)}$
&& $\pi_a\circ \pi'_{b}$   &\cr
}\hrule}}}}
\centerline{
\hbox{{\bf Table 4:}{\it ~~ Chiral spectrum in $d=6$}}}
}
\vskip 0.5cm
\noindent
Because there are no other contributions to the irreducible
Tr$(F_a^4)$ anomaly coefficient these cancel automatically by the tadpole 
cancellation \tadhom. The gravitational $R^4$ anomaly comes out to be 
\eqn\ano{
A_{\rm op}= 14\,  \pi_{{\rm O}7}\circ \pi_{{\rm O}7} ,
}
in terms of ${\cal N}=(0,1)$ supermultiplets.  
In fact, it had to be expected that the net contribution is independent 
of the concrete set of charged matter and only depends on its total 
self-intersection, since 
\ano\ must cancel the contribution $A_{\rm cl}=273-n_{\rm H}-29\,
n_{\rm T}=28(9-n_{\rm T})$ to the anomaly from the gravity multiplet. 
It now follows that 
\eqn\rela{
\pi_{{\rm O}7}\circ \pi_{{\rm O}7}=2(9-n_{\rm T}) =
{1\over 32} \sum_{a,b}{ N_a N_b( \pi_a \circ \pi_b + \pi_a \circ \pi_b')} ,
}
a strong consistency requirement that relates the topology of the 
O7-plane and the number of tensor multiplets in the effective theory. 
In \BlumenhagenWN\ we have indeed given a purely mathematical proof of 
\rela\ that only rested on the sLag nature of Fix$(\o\sigma)$. 

One can demonstrate that the spectrum of table 
4 reproduces essentially all known orbifold models of type IIB 
orientifolds on K3 
\refs{\rgimpol\rgimjo\rbluma\DabholkarKA\rblumb\rdabol
\rbgka-\refPradisi}, although their results are usually obtained after 
lengthy CFT computations and tedious Chan-Paton algebra. In this 
sense, the concept of intersecting branes also offers a technical 
short-cut to produce such supersymmetric orientifold spectra. 
In the following we shall now discuss just one example of a K3-orbifold. 

\subsec{The $T^4/\ZZ_2$ K3-orbifold}

A general K3-orbifold group $\ZZ_N = \{ \Theta, \Theta^2, \, ... \, , 1
\}$ acts crystallographically on $T^{4}$ which we may assume to
be a direct product of two-dimensional tori  $T^2_I,\ I=1,2$. 
The factorization implies a diagonal period matrix $\tau^{IJ}$, 
\eqn\coord{
dz_I = dx_I + \sum_{J=1}^2 \tau^{IJ} dy_J = dx_I + \tau^I dy_I , \quad
z_I \equiv z_I + 1, \quad z_I \equiv z_I + \tau^I , 
}
with diagonal orbifold action 
\eqn\orb{
\Theta z_I = e^{2\pi iv_I } z_I, \quad v_1+v_2 = 0 ,
}
and $\o\sigma$ reflecting $\Im(z_I)$.
The entire orientifold group is generated by $\Omega\o\sigma$ and $\Theta$.
The calibration condition is now rephrased in terms of the relative 
angles $\varphi_I$ of any D7-brane with respect to the O7-plane, 
\eqn\angletheta{
\varphi_1 \pm \varphi_2  = \theta ,
}
with some fixed angle $\theta$ and an arbitrary sign. 
To preserve the same supersymmetry as does the O7-plane, 
one needs to demand 
\eqn\angle{
\varphi_1 + \varphi_2 = 0 .
}
For the sake of brevity we now specialize to give just a single example 
of a K3-orbifold, the 
$\ZZ_2$ orbifold limit of K3. The action of $\ZZ_2$ on the $z_1,z_2$ 
is defined by $v_I=(1/2,-1/2)$ as in \orb.
The homology includes some 2-cycles $\pi_a$ on the K3 which are inherited
from the torus cycles $\o\pi_a$ , corresponding to massless modes in the untwisted 
sector of the CFT in the singular limit.
They are organized in orbits under $\ZZ_2$ 
\eqn\orbit{
\pi_a = \o\pi_a + \Theta \o\pi_a . 
}
The intersection form of these cycles is given by
\eqn\inta{
\pi_a\circ\pi_b={1\over 2} \left(\sum_{i=0}^{1} \Theta^i \o\pi_a
  \right) \circ \left(\sum_{j=0}^{1} \Theta^j \o\pi_b \right) .
}
In the case at hand, there are six elements $\o\pi_{ij}$ in
$H_2(T^4,\ZZ)$, which we denote as
\eqn\basist{ \{\o\pi_{13},\o\pi_{24},\o\pi_{14},\o\pi_{23},
               \o\pi_{12},\o\pi_{34} \} .}
The indices $(1,2,3,4)$ 
are referring to the coordinates $(x_1,y_1,x_2,y_2)$ along the 
four 1-cycles of the $T^2_I$. Their intersection form reads
\eqn\inti{   I_{T^4}=
\left(\matrix{ 0 & 1 \cr 1 & 0 } \right)
\otimes \, {\rm diag}( 1,-1,-1) .
}
Each of these six 2-cycles $\o\pi_{ij}$ on $T^4$ gives rise to a
2-cycle on the orbifold, $\pi_{ij}$. 
A subtlety arises in their proper normalization as generators of 
$H_2(T^4/\ZZ_2,\ZZ)$, which is provided by $\pi_{ij} = 2\o\pi_{ij}$. 
The intersection matrix
\eqn\inti{
I^{\rm Torus}_{T^4/\ZZ_2}= 2\, I_{T^4}
}
follows for the cycles on the orbifold. 

In addition the resolution of the fixed points of $\Theta$ give
rise to exceptional 2-cycles, massless fields
in the twisted sectors of the orbifold CFT. 
For the $\ZZ_2$ orbifold there are 16 2-cycles blown-up to 
$\IC\IP^1$ at the 16 fixed points $P_{ij}$. The 
exceptional divisors are then denoted $e_{ij}$.
Their intersections read
\eqn\intb{
e_{ij}\circ e_{kl}=-2\,\delta_{ik}\delta_{jl}   ,
}
the Cartan matrix of $A_1^{16}$. 
As can be deduced from comparing to the CFT limit 
\refs{\rbgka,\rbgkb\rbgkbsum-\refHonecker}, the
${\rm O}7$-planes only wrap 2-cycles $\pi_a$ inherited from the
torus and no exceptional divisors. Its homology class is then given by 
\eqn\oplane{
\pi_{{\rm O}7}=2( \pi_{13} + \pi_{24} ) .
}
To determine the action of $\Omega\o\sigma$ on the cohomology of K3 
one needs to take the intrinsic reflection of all twisted fields 
by $\Omega$ into account. We then write $[\o\sigma]=(-1)\otimes p$,
with some permutation $p$ of twisted sectors, 
\eqn\actio{
                 e_i \mapsto -e_{p(i)} .
}
The actual operation $p$ depends crucially on the complex structure of the 
background torus, for which two inequivalent choices are compatible with 
the required symmetries. In the particular case we shall be discussing, 
this distinction is unimportant, however. 
Referring to the {\bf A} type choice, the complex structure
$\tau^I$ of any single $T^2_I$ is defined by selecting the two lattice
vectors $1,\ \tau^I = i\Im(\tau^I)$ as a basis, 
where $\o\sigma$ reflects along the real line.
With this choice of complex structure all the fixed points
$P_{ij}$ are geometrically invariant. Through the intrinsic
reflection of the blow-up mode, $[\o\sigma]$ then reflects all
$e_{ij}$. On $T^4$ only the cycles $\o\pi_{13}$ and $\o\pi_{24}$
are invariant and the action of $[\o\sigma]$ is summarized together by
\eqn\inti{   [\o\sigma]_{\bf AA}=
{\rm diag}\left( {\bf 1}_2 , {\bf -1}_{20} \right)
, }
with ${\bf 1}_n$ denoting unit matrices of rank $n$.
From this the number of tensor multiplets follows as the 
number of eigenvalues $+1$ minus 1 to be $n_{\rm T}=1$.
Computing the self-intersection number of the orientifold plane \oplane\
we find $\pi_{{\rm O}7}\circ \pi_{{\rm O}7} =16$, consistent with \rela. 

For the simple case of a $\ZZ_2$ orbifold group, one can directly 
compare the results of this procedure to the standard $\ZZ_2$ orientifold 
of type IIB string theory \refs{\sagn,\rgimpol}, since the
projection by $\Omega\o\sigma$ is equivalent to the standard
projection by $\Omega$ upon performing T-dualities along the two
circles parameterized by $\Im(z_i)$. One particular solution 
of the tadpole constraints will then be found recovering the
spectrum and gauge group first discovered by Bianchi and
Sagnotti and described in terms of D-branes by Gimon and Polchinski. 
In order to do so, we introduce just two
stacks of fractional D7-branes with multiplicities $N_1=N_2=16$,
supporting $U(16)^2$. The cycles are
\eqn\exama{\eqalign{
\pi_1&={1\over 2}\left( \pi_{13} \right)
                  +{1\over 2}\left(e_{11}+e_{12}+e_{21}+e_{22} \right), \cr
                    \pi_2&={1\over 2}\left( \pi_{24} \right)
                  +{1\over 2}\left(e_{11}+e_{13}+e_{31}+e_{33} \right), \cr
}
}
plus their images under $\Omega\o\sigma$. Their intersections follow 
\eqn\intnuma{\eqalign{  
&\pi_1\circ\pi_1=-2, \quad\quad  \pi_2\circ\pi_2=-2,\quad\quad
                       \pi_1\circ\pi'_1=2, \quad\quad  \pi_2\circ\pi'_2=2, \cr
                       &\pi_1\circ\pi_{{\rm O}7}=2, \quad\quad  \pi_2\circ\pi_{{\rm O}7}=2,
         \quad\quad   \pi_1\circ\pi_2=0, \quad\quad  \pi_1\circ\pi'_2=1 \cr}
}
to produce the chiral massless spectrum shown in table 5. 
\vskip 0.8cm
\vbox{
\centerline{\vbox{
\hbox{\vbox{\offinterlineskip
\def\tablespace{height1pt&\omit&&\omit&\cr}
\def\tablerule{\tablespace\noalign{\hrule}\tablespace}

\hrule\halign{&\vrule#&\strut\hskip0.2cm\hfill #\hfill\hskip0.2cm\cr
& Representation && Multiplicity &\cr
\tablerule
& $[({\bf Adj,1})+({\bf 1, Adj})]_{(2,1)}$ && 2 &\cr
& $[({\bf A,1})+({\bf 1, A})+c.c.]_{(1,2)}$ && 2 &\cr
& $[({\bf 16},{\bf 16})+c.c.]_{(1,2)}$ && 1 &\cr
}\hrule}}}}}
\centerline{
\hbox{{\bf Table 5:}{\it ~~ BS/GP model }}}
\vskip 0.5cm
\noindent
which is identical to the results of \refs{\sagn,\rgimpol}.
Note, that in the orbifold limit chirality was induced
by a non-trivial projection on the Chan-Paton indices, while the 
intersections of the brane were vanishing. However, in
the smooth case the gauginos are localized at the
self-intersections of the D7-branes. At the orbifold point the model is
supersymmetric, as here the D7-branes simply lie on top of the
orientifold plane. According to the discussion above, this does not change 
as long as all D7-branes wrap cycles calibrated by $\Re(\Omega_2)$.

In the view of the very general prescriptions to construct intersecting 
brane models, it is apparent that the solution found by the CFT methods 
is not the only supersymmetric vacuum of the $T^4/\ZZ_2$ K3-orbifold. 
To demonstrate this explicitly, let us consider an example that has a 
gauge group of reduced rank. It involves just a single stack of $N=16$
branes, wrapped on
\eqn\exama{
\pi={1\over 2}\left( \pi_{13}+\pi_{24}+\pi_{14}
                   +\pi_{23} \right)
                  +{1\over 2}\left(e_{11}+e_{44}+e_{14}+e_{41} \right) 
}
together with its image stack, with gauge group $U(16)$.
The relevant intersection numbers are 
\eqn\intnum{ \pi\circ\pi=-2, \quad\quad  \pi\circ\pi'=4, \quad\quad
             \pi\circ\pi_{{\rm O}7}=4 ,
}
giving rise to the chiral massless spectrum
\vskip 0.8cm
\vbox{
\centerline{\vbox{
\hbox{\vbox{\offinterlineskip
\def\tablespace{height1pt&\omit&&\omit&\cr}
\def\tablerule{\tablespace\noalign{\hrule}\tablespace}

\hrule\halign{&\vrule#&\strut\hskip0.2cm\hfill #\hfill\hskip0.2cm\cr
& Representation && Multiplicity &\cr
\tablerule
& $[{\bf Adj}]_{(2,1)}$ && 2 &\cr
& $[{\bf A+\o{A}}]_{(1,2)}$ && 4 &\cr
}\hrule}}}}
\centerline{
\hbox{{\bf Table 6:}{\it ~~ Chiral spectrum }}}}
\vskip 0.5cm
\noindent
Now the tadpoles are not canceled locally even at the singular 
orbifold point before blowing-up. Therefore, supersymmetry is not 
preserved in a trivial manner and the vanishing of the scalar D-term 
potential imposes constraints on the complex structure moduli 
of the torus. The scalar potential
\susy\ at the orbifold point reduces to 
\eqn\tadp{
{\cal V} = T_7 e^{-\phi_6}\left[
\prod_{I=1}^2\sqrt{\Im(\tau^I)+{1\over \Im(\tau^I)}}-
\left(\sqrt{\Im(\tau^1)\, \Im(\tau^2)}+{1\over
              \sqrt{\Im(\tau^1)\, \Im(\tau^2)}}\right) \right] }
which vanishes precisely if $\Im(\tau^1)=\Im(\tau^2)$. 
Moving away from this supersymmetric 
locus, the intersection angles do no longer satisfy \angle,
supersymmetry is broken spontaneously, and an open string tachyon appears.  
The corresponding FI term depends on 
$\xi \sim \Im(\tau^1)-\Im(\tau^2)$.
In six dimensions the D-term potential has the general form
\eqn\dtermb{
{\cal V}_{\rm D-term} \sim
\biggl(\sum_i q^i |\phi_i|^2- \sum_i q^i |\widetilde\phi_i|^2-\xi\biggr)^2\, ,
}
where now $\phi_i$ and $\widetilde\phi_i$ denote two complex scalars
with charge $q^i$. Independent of
the sign of $\xi$, one always gets a tachyonic mode if $\xi\ne 0$,
which is in accord with the string theory picture.

\newsec{M-theory lift of ${\cal N}=1$ intersecting brane worlds via
$G_2$ manifolds}

As it is well known, 11-dimensional M-theory is supposed to describe
the strong coupling regime of the type IIA superstring; hence we
expect that by lifting intersecting brane world models with
non-abelian gauge groups and chiral fermions interesting
non-perturbative informations about the gauge dynamics can be
obtained. Furthermore, as we will describe, an ${\cal N}=1$
supersymmetric intersecting brane world scenario with only D6-branes,
and possibly O6-planes, can be nicely described in purely geometrical
terms via M-theory compactification on a seven-dimensional Ricci-flat
manifold $X_7$ with reduced $G_2$ holonomy group (for some papers on
M-theory compactifications on $G_2$ spaces see
\refs{\PapadopoulosDA\AcharyaGB\AtiyahZZ\CveticYA\BrandhuberYI
\AtiyahQF\rwitten\CveticZX\rwa\BlumenhagenJB\RoibanCP\EguchiIP
\CurioDZ\refTamarFriedmann\BrandhuberKQ\rberbra\BehrndtYE
\AnguelovaDD-\BehrndtXM,\rcveticb}).
This can be easily seen by looking at the fields which couple to the
6-brane background. The D6-branes are the magnetically charged
monopoles of the KK vector, and thus only couple to components of the
eleven-dimensional metric, the RR 1-form $C_1$ and the dilaton of type
IIA.  The geometric lift of isolated D6-branes in flat ten-dimensional
space-time is then a non-trivial $U(1)$ fibration over an
$S^2\subset\IR^3$, a Taub-Nut space.  Similarly, the O6-planes lift to
an Atiyah-Hitchin space.

More precisely,
the relation between the geometrical M-theory picture and the type IIA
brane picture is as follows: 
If $X_7$ has a suitable $U(1)$ isometry, one obtains a type IIA
superstring interpretation upon dimensional reduction to ten
dimensions. This circle is usually non-trivially fibered over a
six-dimensional base $B_6$ which serves as the geometric background of
the corresponding IIA superstring theory, i.e.
$B_6=X_7/U(1)$. The space $B_6$ is in general
non Ricci-flat, whereas the curvature of $B_6$ reflects the gravitational
back reaction of the 6-branes on the type IIA metric. The
specific form of the IIA brane configuration depends very much on the
choice of the $U(1)$ action. 
In order to obtain a configuration that contains
D6-branes, one has to ensure that the $U(1)$ action has a codimension
4 fixed point set $L$ which describes
the world volume locus of the 6-branes.
In M-theory language non-abelian
gauge bosons arise, if $X_7$ has an A-D-E singularity of codimension
four. The non-abelian gauge bosons correspond to massless M2-branes
wrapped around collapsing 2-cycles. Product gauge groups 
with chiral bi-fundamental matter representations are
provided by colliding singularities, i.e. by two or more
sets of fixed points
$L=L_1 \cup L_2\dots\cup L_i$, which intersect at a point on $X_7$.
In the IIA brane picture this is described by the
intersection of 6-branes. Hence 
massless fermions are supported
by isolated (conical) singularities of codimension 7 of $X_7$,
whose metrics are given in terms of a radial cone on a six-dimensional
base space $Y_6$:
\eqn\conemetric{
ds^2_{X_7}=dr^2+r^2d\Omega^2_{Y_6}\, .}
In order for $X_7$ to have $G_2$ holonomy, it is known that $Y_6$
has to have weak $SU(3)$ holonomy. Resolving the point like singularity at
$r=0$ means that the corresponding
product gauge group gets spontaneously broken to a diagonal subgroup by the
vev of a bifundamental scalar field and that the associated
fermions become massive.

So far, metrics for compact $G_2$ spaces have not yet been 
constructed. However a few examples of non-compact $G_2$ metrics are
explicitly known. One can view these non-compact $G_2$ spaces as
describing the local neighborhood of a compact $G_2$ space around some
local singularity, e.g. around the locus of (intersecting) D6-branes.
Being on a non-compact $G_2$ manifold, the gravitational degrees of
freedom decouple, and one is left only with the local gauge degrees
of freedom. For the corresponding IIA superstring theory this means
that the global RR tadpole conditions do not need to be satisfied, since
part of  RR fluxes can escape to infinity on a non-compact direction.

Basically the known examples of non-compact $G_2$ spaces group
together into two classes \refs{\rbs,\GibbonsER}: one is topologically a
$\IR^4$ bundle over $S^3$ and the other a $\IR^3$
bundle over a quaternionic base space $Q$.
The first class can be e.g. generalized by an $\IR^4/\ZZ_N$
bundle over $S^3$, see \rbs. This situation corresponds
to $N$ wrapped, but non-intersecting
D6-branes around the $S^3$ of the deformed
conifold; the associated gauge theory is 
${\cal N}=1$ $SU(N)$ super Yang-Mills
without chiral matter fields.

In the second class one indeed
obtains examples with intersecting
D6-branes. Specifically, examples with known metrics are given by the 
quaternionic spaces $Q=S^4$ and $Q={\IC\IP}^2$.
Consider briefly the
first example with $Q=S^4$ where the metric of $X_7$ can be written as
a cone on $Y_6={\IC\IP}^3$ \AtiyahQF. The associated fixed point set of the
$U(1)$ action is given by $L={\IR}^3\cup 
{\IR}^3$, meeting at the origin in
$\IR^6$. This corresponds to two intersecting D6-branes at
special angles such that supersymmetry is preserved.
The related field theory is given by an 
${\cal N}=1$, abelian $U(1)\times U(1)$
gauge theory with one charged chiral matter field.

Next let us discuss in some more detail the case of three intersecting 
D6-branes which belongs to $Q={\IC\IP}^2$ \AtiyahQF. The associated 
metric describes a cone on $Y_6=SU(3)/(U(1)\times U(1))$.
Here the fixed point set $L={\IR}^3\cup{\IR}^3\cup{\IR}^3$
corresponds to three D6-branes which intersect in one point at supersymmetric
angles. Resolving the point like singularity at the origin $r=0$ such
that the cone is deformed to a smooth $G_2$ manifold, the fix point
set becomes $L=S^2\times{\IR}\cup{\IR}^3$
with zero intersection of the two branches. This corresponds
to two disjoint D6-branes which do not intersect anymore. Hence there
are no more massless fermions on the resolved singularity.
As explained  in \AtiyahQF\ this has the following nice field
theory interpretation. In the singular case the gauge
group of the three intersecting D6-branes is $U(1)^3$ with three
chiral matter fields $\Phi_1=(1,-1,0)$, $\Phi_2=(-1,0,1)$
and $\Phi_3=(0,1,-1)$. In addition there is also a world sheet instanton
generated
superpotential of the form $W\sim\Phi_1\Phi_2\Phi_3$. The flat directions
of this superpotential always allow one charged scalar field, say
$\phi_3$, to take an arbitrary vev. So in the generic case the gauge group is
Higgsed to $U(1)^2$ and all fermions are massive. This Higgsing
precisely corresponds to resolving the cone singularity. Two
D6-branes recombine into a single D6-brane under the Higgsing, which does
not anymore intersect the third D6-brane.

Now it would be an interesting question how to generalize this
scenario to the case of several intersecting D6-branes with
associated non-abelian gauge structure. In \AtiyahQF\ it was proposed
to consider an ${\IR}^3$ bundle over ${\bf W\IC\IP}^2_{N_1,N_2,N_3}$
with at least two of the indices $N_i$ are equal, say $N_2=N_3$.
This should then correspond to the intersection of 3 stacks of D6-branes.
Just like in the abelian case,
the corresponding gauge group $U(N_1)\times U(N_2)\times U(N_2)$
will then be Higgsed to $U(N_1)\times U(N_2)$ by the vev of the bifundamental
scalar field when resolving the singular cone. In the following we
will describe the work of
\BehrndtXM, where 
it was tried to provide an explicit $G_2$ metric for this set-up
by replacing the homogeneous, quaternionic spaces $Q$, considered so far,
by an non-homogeneous quaternionic space with only two isometries
(see also  \AnguelovaDD\ for related work).
Specifically the quaternionic spaces used in \BehrndtXM\ are based 
on the four-dimensional Minkowskian spaces with anti-self dual Weyl tensor
introduced by Demiansky and Plebanski \PlebanskiGY. The corresponding Euclidean
metric reads
\eqn\quatmetric{ 
ds^2_4 = { p^2 - q^2 \over P} dp^2
+ {p^2 - q^2 \over Q}dq^2   + {P \over p^2 -q^2} \Big(d \tau + q^2
d \sigma \Big)^2 + {Q \over p^2 - q^2} \Big(d \tau + p^2
d \sigma \Big)^2\, ,}
with the forth order polynomials in the two coordinates $p$ and $q$;
\eqn\pol{\eqalign{
P &= - \kappa (p-r_1)(p-r_2)(p-r_3)(p-r_4) \ ,\cr
Q &=  \kappa (q-r_1)(q-r_2)(q-r_3)(q-r_4) \ ,\cr
0 &= r_1 + r_2 +r_3 +r_4 \ .}}
Via the above constraint
the quaternionic space depends on three parameters $r_1$, $r_2$ and
$r_3$. 
The associated 7-dimensional metric with $G_2$ holonomy is given by
\eqn\gtwometric{
ds^2 = {1 \over \sqrt{2 \kappa |u|^2 + u_0}}\,
\big( du^i + \epsilon^{ijk} A^j u^k \big)^2
+ \sqrt{2 \kappa |u|^2 + u_0} \; ds^2_4 \ ,}
which is topologically a ${\IR}^3$ bundle (related to the
coordinates $u^i$) over the quaternionic base space, given by the metric
$ds^2_4$ with  the $SU(2)$ connection $A^i$.
For $u_0\neq$ this space is smooth; however setting $u_0=0$ it
develops a point like singularity, i.e. it will become a cone on
a six-dimensional base $Y_6$.

In order to reduce to the ten-dimensional type IIA string with intersecting
D6-branes, one has to choose an appropriate $U(1)$ Killing vector, which 
corresponds to the 11th M-theory directions. It will be a specific linear
combination
\eqn\killing{
k = \beta_1 \partial_\tau - \beta_2 \partial_\sigma\, .
}
Then the brane locations will depend on the fixed point
set of $k$. Consider the following Killing vector
\eqn\kill{
k =r_3^2 \, \partial_\tau - \partial_\sigma\, .}
As discussed in \BehrndtXM\ there are now two sets of 6-branes located at
\eqn\twodsix{\eqalign{D6_1 :~ &p = r_3 , \ u_1 = u_3 = 0 \ ,\cr
D6_2 :~ &q = r_3 , \ u_1 = u_3 = 0\, }}
But by keeping generic values of the roots, there will be further
codimension 6 fixed points at $q=r_2$, $p=r_4$, $u_1=u_2=0$ and at
$p=r_2$, $q=r_4$, $u_1=u_2=0$. In order to avoid these fixed points we
will set $r_1 = r_2$, which essentially moves these fixed points to
infinity since the metric  develops an infinite throat at
$p \rightarrow r_2=r_1$.
In addition the parameters have to obey the constraint 
$0= 2r_2+ r_3 + r_4$, such that the metric depends on two parameters, say
$r_3$ and $r_4$.
The number of D6-branes at the two fixed point sets is related to the
surface gravity $|\nabla k|$ of the corresponding fixed point set, namely
$N_i \sim {1 \over |\nabla k|_i}$.
Calculating the surface gravity for the
fixed point set given in eq.\twodsix\ gives
\eqn\surface{
|\nabla k|_{1} =|\nabla k|_2 = {\kappa  \over 4} \, (3 r_3+r_4)^2 
(r_4-r_3)\, .}
(That both numbers coincide, is a consequence of the symmetry $p
\leftrightarrow q$ of the metric.)
Since this solution is characterized by two parameters $r_3$ and $r_4$,
it is very tempting to identify this space as the one related to the
weighted projective space ${\bf W\IC\IP}^2_{N_1,N_2,N_2}$. 
In our case the number of 6-branes in two stacks agree
and we expect a gauge group $SU(N_1)^3 $, where
in the deformed case the Higgsing should be done in a way that the
product of two equal gauge groups survives, because the two components
of the fixed point set are related to the same number of 6-branes.  At
the moment, these conclusions are more speculative and further
investigations are necessary.

\vskip3cm

\centerline{{\bf Acknowledgments}}\pano
We would like thank the organizers of SUSY02, {\it The 10th 
International Conference on Supersymmetry and Unification 
of Fundamental Interactions} at DESY Hamburg, of {\it The First International
Conference on String Phenomenology} at Oxford, of Strings 2002 at 
Cambridge, and finally of the {\it 35th International Symposium 
Ahrenshoop on the Theory of Elementary Particles, Recent Developments in 
String/M-Theory and Field Theory} at Berlin. The material presented 
in this article is an extended and combined version of the 
talks given at these conferences. D.L. like to thank Klaus Behrndt,
Gianguido Dall'Agata and Swanpa Mahaptra for the pleasant collaboration on
the material presented in the last chapter. 
The work is supported in part by the EC under the RTN project
HPRN-CT-2000-00131. 

\vfill\eject

\listrefs

\bye
\end